\documentclass[aps,prb,twocolumn,amsmath,amssymb,nofootinbib,superscriptaddress,floatfix,eqsecnum]{revtex4-1}

\usepackage{amsmath}
\usepackage{amssymb}
\usepackage{amsthm}
\usepackage[dvipdf]{color}
\usepackage{graphicx}
\usepackage{dcolumn}
\usepackage{bm} 
\newcommand*{\rom}[1]{\expandafter\romannumeral #1}

\begin{document}
\title{Real Space Renormalization: A Generic Microscopic Theory for Low-Temperature Avalanches in Static Strained Insulating Glass}

  \author{Di Zhou}

 \affiliation{
 Department of Physics,
  University of Illinois at Urbana-Champaign,
  1110 West Green St, Urbana,
   Illinois 61801, USA
 }

\date{\today}

\begin{abstract}
We propose a microscopic model to study the avalanche problem of insulating glass deformed by external static uniform strain below $T=60$K. We use three-dimensional real-space renormalization procedure to carry out the glass mechanical susceptibility at macroscopic length scale. We prove the existence of irreversible stress drops in amorphous materials, corresponding to the steep positive-negative transitions in glass mechanical susceptibility. We also obtain the strain directions in which the glass system is brittle. The irreversible stress drops in glass essentially come from non-elastic stress-stress interaction which is generated by virtual phonon exchange process.
\end{abstract}

\maketitle

\section{Introduction}
Amorphous solids (Glasses), which are known to have non-crystalline structure like fluids, possess solid-like behaviors strikingly different from crystalline solids such as saturation\cite{Graebner1983}, universality of internal friction\cite{Pohl2002}, linear heat capacity\cite{Zeller1971} etc.. One of these properties is called avalanche phenomena, in which the glass stress-strain curve presents a steep drop to a lower value at certain critical external strain. To explain these universal properties, Anderson, Halperin and Varma\cite{Anderson1972} group and Phillips\cite{Phillips1987} independently proposed a phenomenological model known as tunneling-two-level-system (TTLS). It not only explained several of existing experimental observations, but also predict new phenomena such as phonon echo\cite{Golding1976}. However, at least to the author's knowledge, there was no direct explaination to the mechanical avalanche problem in glass according to the original TTLS model papers\cite{Anderson1972, Phillips1987}.

The glass mechanical avalanche phenomena is referred to a broad distribution of irreversible (inelastic) stress drops in non-quantum, amorphous materials undergoing quasi-static external strain\cite{Langer1998, Dahmen2009}. One of the experiments on the glass fracture behavior was in silica glasses and polymers\cite{Fineberg1991} at room temperatures. Based on two-dimensional molecular dynamics simulations, M. L. Falk and J. S. Langer\cite{Langer1998} developed Shear-Transformation-Zone (STZ) model to explain low-temperature shear deformations in metallic glasses. 

The purpose of this paper is to develop a microscopic field theory model, namely the generic coupled block model, to investigate the mechanical avalanche phenomena in three-dimensional insulating glass. More specifically, we want to prove the ``existence of inelastic stress drops" in quantum amorphous materials under the deformation of external static, uniform strain. Consider an external static strain $\bm{e}$. The amorphous material will provide a corresponding stress response $\bm{T}(\bm{e})$. Since we want to prove the ``existence of steep stress drops in stress-strain relation", it is more convenient to prove the ``existence of positive-negative transitions in glass mechanical susceptibility-strain relation", where the glass mechanical susceptibility is defined as the derivative of stress with respect to strain: $\bm{\chi}(\bm{e})=\delta\bm{T}(\bm{e})/\delta \bm{e}$. Therefore, the main purpose of this paper is actually to investigate the ``existence of positive-negative transitions in glass mechanical susceptibility $\bm{\chi}(\bm{e})$" when external strain exceeds certain critical values, $\bm{e}_{\rm crit}$.


The reader should be aware that it is the first time to apply our ``generic coupled block model" in glass mechanical avalanche problems. Therefore our purpose is not to solve the entire avalanche problems from microscopic point of view; instead we want to provide some first-step results for future people to continue studying this problem from our model.

The set up of our model begins from the generalization of tunneling-two-level-system (TTLS) model to generic multiple-level-system (MLS) model. Similar as TTLS model, phonon strain field could couple to multiple-level-systems (MLSs). As the coupling with phonon strain field, multiple-level-systems (MLSs) must generate a mutual RKKY-type interaction\cite{Joffrin1976} due to virtual phonon exchange process. Finally, our glass Hamiltonian is the summation of long-wavelength phonon contribution, a set of MLSs, the coupling between MLS and phonon strain field, and the mutual RKKY-type interaction between MLSs. Since we do not take conducting electrons into consideration, the model only applys for insulating glasses. Further considerations regarding conducting electron Hamiltonian, electron-phonon coupling and electron-MLS coupling are required to explore the ductility of metallic glass.

We want to use this model to carry out macroscopic length scale glass mechanical susceptibility from microscopic length scale. The effective starting microscopic length scale is of order $\sim 50\AA$, corresponding to the characteristic thermal phonon wavelength with the temperature of order $60$K. Since the thermal phonon wavelength can be no smaller than the starting microscopic length scale, our theory is only valid below 60K. We will discuss the reasoning in details in section 3. However, at least to the author's knowledge, all of glass avalanche experiments are taken under room temperatures or glass transition temperatures\cite{Argon2005, Gauthier2004, Dahmen2014, Fineberg1991, Fineberg1992} ($T\sim 300$K). We hope more glass avalanche experiments could be taken below 60K to test the validity of our model.

The paper is organized as follows: in section 2 we first generalize glass two-level-system model to multiple-level-system model. Then we give a detailed derivation of our generic coupled block model by introducing the mutual RKKY-type interaction between multiple-level-systems. We introduce the most important concept of this paper, namely non-elastic stress-stress susceptibility at the end of section 2(A). In section 3 we treat RKKY-type interaction as a perturbation, to set up the relation between small and large length scale non-elastic stress-stress susceptibilities. By repeating such recursion relation, we eventually carry out non-elastic stress-stress susceptibility at macroscopic length scale. In section 4 we prove the existence of positive-negative transitions in non-elastic stress-stress susceptibility with the increase of external static strain. We further prove the existence of positive-negative transitions in glass total mechanical susceptibility.

\section{The Model}

\subsection{The Set up of Problem}
We consider a block of amorphous material under the deformation of static, uniform strain. With the slowly increasing strain the bulk glass behaves elastically until it reaches critical strain value. The stress ($\bm{T}$) v.s. strain ($\bm{e}$) curve shows a steep drop. Let us consider the glass mechanical susceptibility, defined as follows: ${\chi}_{ijkl}(\bm{e})=\delta {T}_{ij}(\bm{e})/\delta e_{kl}$. At critical strain field when irreversable process happens, glass mechanical susceptibility presents an abrupt positive-negative transition. In this paper our main goal is to prove the existence of such positive-negative transitions in mechanical susceptibility, which is shown in Fig.1 as follows:
\begin{figure}[hp]
\centering
\includegraphics[scale=0.27]{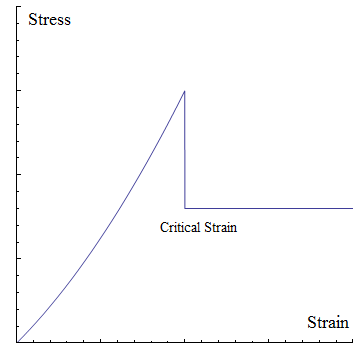}
\quad
\includegraphics[scale=0.27]{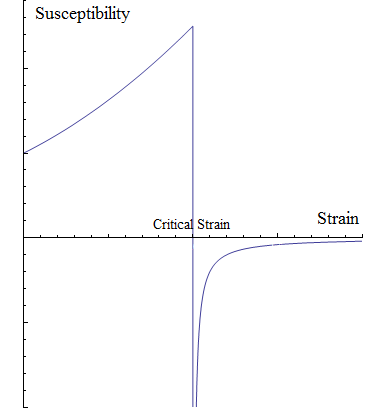}
\caption{As an illustration of stress-strain curve, the left picture shows a steep drop of stress. As an illustration of susceptibility-strain curve, the right picture shows a positive-negative susceptibility transition, where the mechanical susceptibility is the first order derivative of stress with respect to external strain field.}
\end{figure}

Our purpose is to prove the existence of such positive-negative transition of mechanical susceptibility in glass materials. To explore this problem, we begin our discussion from the famous tunneling-two-level-system model (TTLS model)\cite{Phillips1987}. In this model we assume that there are a group of TTLSs randomly embedded in the glass material, with the location $\vec x_i$ for the $i$-th TTLS. The effective glass Hamiltonian $\hat{H}^{\rm tot}$ in TTLS theory is the summation of two-level-system Hamiltonian, long wavelength phonon Hamiltonian, and the coupling between TTLS and strain field (phonon field):
\begin{eqnarray}\label{0.1}
\hat{H}^{\rm tot} & = & \hat{H}^{\rm el}+\sum_i\frac{1}{2}\left(
\begin{array}{cc}
E_i & 0\\
0 & -E_i\\
\end{array}
\right)\nonumber \\
 & {} & +\frac{1}{2}\sum_i \left(
\begin{array}{cc}
D_i & M_i\\
M_i & -D_i\\
\end{array}
\right)\cdot\bm{e}(\vec x_i)
\end{eqnarray}
where the first, second and third terms stand for long wavelength phonon Hamiltonian (we will also call it ``purely elastic Hamiltonian $\hat{H}^{\rm el}$"), the Hamiltonian of a group of two-level-systems, and the coupling between every two-level-system and phonon strain field at corresponding position $\vec x_i$, respectively. The two-level-system Hamiltonian is written in the representation of energy eigenvalue basis, with $E_i=\sqrt{\Delta_i^2+\Delta_{0i}^2}$ the energy splitting; $D_i=\Delta_i/E_i$ and $M_i=\Delta_{0i}/E_i$ are diagonal and off-diagonal matrix elements of the coupling between two-level-system and strain field, and by definition they are no greater than 1; $\bm{e}(\vec x_i)$ is the local intrinsic strain field at the position of the $i$-th two-level-system.

The purpose of this subsection is to develop the multiple-level-system from the generalization of 2-level-system model. At this moment, we have not applied any external strain field yet. We will consider external strain in subsection 2(C). We begin our model by considering a single block of glass with the length scale $L$ much greater than the atomic distance $a\sim 10\AA$. In the next subsection 2(B), we will combine a group of such single blocks to form a ``super block". We will consider the RKKY-type interaction between these single blocks, which is generated by virtual phonon exchange process. For now, we do not consider RKKY-type interaction and focus only on the Hamiltonian of single block glass.

We further define intrinsic strain field $e_{ij}(\vec x)$ at position $\vec x$: if $\vec u(\vec x)$ denotes the displacement relative
to some arbitrary reference frame of the matter at point $\vec x$, then strain field is defined as follows
\begin{eqnarray}\label{1}
e_{ij}(\vec x)=\frac{1}{2}\left(\frac{\partial u_i(\vec x)}{\partial x_j}+\frac{\partial u_j(\vec x)}{\partial x_i}\right)
\end{eqnarray}
We write down the general Hamiltonian of glass as $\hat{H}^{\rm tot}$. Let us separate out from the glass general Hamiltonian $\hat{H}^{\rm tot}$, the purely elastic contribution $\hat{H}^{\rm el}$. It can be represented either by phonon creation-annihilation operators or strain fields:
\begin{eqnarray}\label{6}
 & {} & \hat{H}^{\rm el}
 = \sum_{k\alpha}\hbar\omega_{k\alpha}\left(\hat{a}_{k\alpha}^{\dag}\hat{a}_{k\alpha}+\frac{1}{2}\right)
 = {\rm Const}+\nonumber \\
 & {} & \frac{1}{2}\int d^3x\left(\sum_{ijkl}\chi_{ijkl}^{\rm el}e_{ij}(\vec x)e_{kl}(\vec x)+\sum_i\rho \dot{u}_i^2(\vec x)\right)
\end{eqnarray}
where $\alpha=l,t$ represents phonon polarization, i.e., longitudinal and transverse phonons. We define the elastic stress tensor $\hat{T}_{ij}^{\rm el}(\vec x)$ by 
\begin{eqnarray}\label{6.1}
\hat{T}_{ij}^{\rm el}(\vec x)=\frac{\delta \hat{H}^{\rm el}}{\delta e_{ij}(\vec x)}
\end{eqnarray}
From the elastic Hamiltonian Eq.(\ref{6}), the elastic stress tensor is further given by $
\hat{T}_{ij}^{\rm el}(\vec x)=\sum_{kl}\chi_{ijkl}^{\rm el}e_{kl}(\vec x)
$, where $\chi_{ijkl}^{\rm el}$ is the purely elastic part of susceptibility. Since by definition the properties of an isotropic amorphous material must be invariant under real space SO(3) rotational group, symmetry considerations require $\chi_{ijkl}^{\rm el}$ to have the generic form\cite{Zeller1971}
\begin{eqnarray}\label{7}
{\chi}^{\rm el}_{ijkl}=\left(\rho c_l^2-2\rho c_t^2\right)\delta_{ij}\delta_{kl}+\rho c_t^2\left(\delta_{ik}\delta_{jl}+\delta_{il}\delta_{jk}\right)\quad
\end{eqnarray}
where $\rho$ is the mass density and $c_{l,t}$ is the longitudinal/transverse sound velocity.

Subtracting the purely elastic Hamiltonian $\hat{H}^{\rm el}$, we name the left-over glass Hamiltonian $(\hat{H}^{\rm tot}-\hat{H}^{\rm el})$ as ``the non-elastic part of glass Hamiltonian, $\hat{H}^{\rm non}$". We expand the left-over Hamiltonian $\hat{H}^{\rm non}$ up to the first order expansion of long wavelength intrinsic phonon strain field. We name the coefficient of the first order expansion to be ``non-elastic stress tensor $\hat{T}_{ij}^{\rm non}(\vec x)$", defined as follows:
\begin{eqnarray}\label{9}
 & {} & \hat{H}^{\rm non} = \hat{H}^{\rm tot}-\hat{H}^{\rm el}\nonumber \\
 & {} & \hat{H}^{\rm non} = \hat{H}^{\rm non}_0+\int d^3x\sum_{ij}e_{ij}(\vec x)\hat{T}^{\rm non}_{ij}(\vec x)+\mathcal{O}(e_{ij}^2)\nonumber \\
 & {} & \hat{T}_{ij}^{\rm non}(\vec x) = \frac{\delta \hat{H}^{\rm non}}{\delta e_{ij}(\vec x)}
\end{eqnarray}
Now let us stop for a moment and compare Eq.(\ref{9}) with Eq.(\ref{0.1}): $\hat{H}^{\rm tot}$ and $\hat{H}^{\rm el}$ in Eq.(\ref{9}) corresponds to the glass total Hamiltonian and purely elastic Hamiltonian in Eq.(\ref{0.1}) of TTLS model, respectively; the zeroth order expansion of non-elastic Hamiltonian $\hat{H}^{\rm non}$ with respect to strain field $e_{ij}$, $\hat{H}^{\rm non}_0$, is the generalization from two-level-system Hamiltonian to multiple-level-system Hamiltonian; non-elastic stress tensor $\hat{T}_{ij}^{\rm non}$ is the multiple-level generalization of the term which couples to strain field in TTLS model. In the rest of this paper, we denote $\hat{H}^{\rm non}_0$ to be the non-elastic Hamiltonian excluding the coupling between intrinsic phonon strain $e_{ij}(\vec x)$ and non-elastic stress tensor $\hat{T}_{ij}^{\rm non}(\vec x)$. We denote $\hat{H}^{\rm non}$ to be the non-elastic Hamiltonian including the stress tensor--intrinsic phonon strain coupling (see the second equation of Eq.(\ref{9})).

Let us denote $|m\rangle$ and $E_m$ to be the $m$-th eigenstate and eigenvalue of the non-elastic Hamiltonian $\hat{H}_0^{\rm non}$. Such set of eigenbasis $|m\rangle$ is a generic multiple-level-system. Note that in general, in the representation in which $H_0^{\rm non}$ is diagonal, $\hat{T}_{ij}^{\rm non}$ will have both diagonal and off-diagonal matrix elements.

Now we can define the most important quantity of this paper, namely the non-elastic stress-stress susceptibility (i.e., linear response function). Consider an external infinitesimal sinusoidal strain field, $e_{ij}(\vec x, t)=e_{ij}(e^{i(\vec k\cdot \vec x-\omega t)}+e^{-i(\vec k\cdot \vec x-\omega t)})$, where $e_{ij}$ is real. The non-elastic Hamiltonian $\hat{H}^{\rm non}$ will provide a corresponding response $\langle \hat{T}_{ij}^{\rm non}\rangle(\vec x, t)=(\langle \hat{T}_{ij}^{\rm non}\rangle e^{i(\vec k\cdot \vec x-\omega t)}+c.c)$, where $\langle \hat{T}_{ij}^{\rm non}\rangle$ is in general complex. Then we can define the complex response function\cite{Anderson1986} (non-elastic stress-stress susceptibility) $\chi_{ijkl}^{\rm non}(\vec k, \omega)$
\begin{eqnarray}\label{4}
\chi_{ijkl}^{\rm non}(\vec k, \omega) & = & \frac{\delta  \langle\hat{T}^{\rm non}_{ij}\rangle}{\delta e_{kl}}(\vec k,\omega)
\end{eqnarray}
In the rest of this paper we will always use $\hat{H}$, $\hat{H}_0$, $\chi_{ijkl}$ and $\hat{T}_{ij}$ to represent non-elastic Hamiltonians $\hat{H}^{\rm non}$, $\hat{H}_0^{\rm non}$, susceptibility $\chi_{ijkl}^{\rm non}$ and stress tensor $\hat{T}^{\rm non}_{ij}$ respectively, while we use $\hat{H}^{\rm el}$, $\chi_{ijkl}^{\rm el}$ and $\hat{T}_{ij}^{\rm el}$ to represent the elastic Hamiltonian, susceptibility and stress tensor, respectively.

In Eq.(\ref{4}) the ``average" of non-elastic stress tensor operator, $\langle \hat{T}_{ij}\rangle(\vec x, t)$, is defined as follows: (please note from now on we use $\hat{T}_{ij}$ to stand for $\hat{T}_{ij}^{\rm non}$)
\begin{eqnarray}\label{100}
\langle \hat{T}_{ij}\rangle(\vec x, t)=\sum_m\frac{e^{-\beta E_m}}{\mathcal{Z}}\langle m,t|\hat{T}_{ij}(\vec x)|m,t\rangle\quad
\end{eqnarray}
where $|m, t\rangle$ is the $m$-th eigenstate wave function of $\hat{H}_0$ perturbed by externally induced infinitesimal perturbation $\int d^3x\,e_{ij}(\vec x, t)\hat{T}_{ij}(\vec x)$, and $\mathcal{Z}$ is the partition function $\mathcal{Z}=\sum_me^{-\beta E_m}$. In the rest of this paper, we use ``non-elastic susceptibility" to short for ``non-elastic stress-stress susceptibility". Combining the definitions Eq.(\ref{4}) and Eq.(\ref{100}), we see non-elastic susceptibility is the function of temperature. However, for notational simplicity we write $\chi_{ijkl}(\vec k,\omega; T)$ as $\chi_{ijkl}(\vec k,\omega)$ in the rest of this paper. By using linear response theory, we expand $\langle \hat{T}_{ij}\rangle(\vec x, t)$ up to the first order of perturbation $\int d^3x\,e_{ij}(\vec x, t)\hat{T}_{ij}(\vec x)$, to calculate non-elastic susceptibility Eq.(\ref{4}) with the following process:

We use the same language as TTLS model, that the non-elastic susceptibility can be expressed in the relxation and resonance parts. The relaxation susceptibility comes from the energy eigenvalue shift due to the diagonal matrix elements of perturbation $e_{ij}(\vec x, t)\hat{T}_{ij}$, while the resonance susceptibility comes from the off-diagonal matrix elements of it. We use $\chi_{ijkl}^{\rm rel}(\vec x, \vec x'; \omega)$ and $\chi_{ijkl}^{\rm res}(\vec x, \vec x'; \omega)$ to stand for the relaxation and resonance susceptibilities respectively. Let us denote $\tau$ to be the effective thermal relaxation time for the glass single block Hamiltonian at temperature $T$. The non-elastic susceptibility is therefore expressed as follows,
\widetext
\begin{eqnarray}\label{11.1}
\chi_{ijkl}(\vec x, \vec x'; \omega) & = &  {\chi_{ijkl}^{\rm rel}(\vec x, \vec x'; \omega)}+ \chi_{ijkl}^{\rm res}(\vec x, \vec x'; \omega)\nonumber \\
\chi_{ijkl}^{\rm rel}(\vec x, \vec x'; \omega) & = & \frac{\beta}{{1-i\omega\tau}}\bigg(\sum_{nm}P_nP_m\langle n|\hat{T}_{ij}(\vec x)|n\rangle\langle m|\hat{T}_{kl}(\vec x')|m\rangle-\sum_n P_n\langle n|\hat{T}_{ij}(\vec x)|n\rangle \langle n|\hat{T}_{kl}(\vec x')|n\rangle\bigg)\nonumber \\
\chi_{ijkl}^{\rm res}(\vec x, \vec x'; \omega) & = & -\frac{1}{\hbar}\sum_{n}\sum_{m\neq n}P_m\frac{\langle n|\hat{T}_{ij}(\vec x)|m\rangle\langle m|\hat{T}_{kl}(\vec x')|n\rangle}{\omega+(E_n-E_m)/\hbar+i\eta}
+
\frac{1}{\hbar}\sum_{m}\sum_{n\neq m}P_n\frac{\langle n|\hat{T}_{ij}(\vec x)|m\rangle\langle m|\hat{T}_{kl}(\vec x')|n\rangle}{\omega+(E_n-E_m)/\hbar+i\eta}
\end{eqnarray}
\endwidetext
where $P_n=e^{-\beta E_n}/\mathcal{Z}$ is the $n$-th eigenstate probability function, $\mathcal{Z}$ is the partition function, $\beta$ is the inverse of temperature $\beta=(k_BT)^{-1}$, and $\eta$ is a phenomenological parameter to represent the higher order corrections of non-elastic susceptibility due to the coupling between intrinsic strain field and non-elastic stress tensor: $\sum_{ij}e_{ij}(\vec x)\hat{T}_{ij}(\vec x)$.

Please note it is an approximation that we use the parameter $\tau$ to represent the effective thermal relaxation time of glass non-elastic part of Hamiltonian $\hat{H}_0$: in principle, the relaxation process of the $n$-th state is the summation of all relaxation processes between $m$-th state and $n$-th state, $\forall m\neq n$. The effective thermal relaxation time $\tau_n$ is different for different quantum number $n$. Generally speaking, one cannot use a simple parameter $\tau$ to stand for the thermal relaxation process for an arbitrary multiple-level-system. However, in this paper we want to discuss the mechanical properties of amorphous materials under the deformation of external static strain with $\omega=0$. Therefore, we always have the important relation, $\omega\tau_n=0$, for arbitrary quantum number $\forall n=0,1,2,\ldots$. In the special case of static external strain, using a simple parameter $\tau$ to represent the thermal relaxation process does not harm our theory. In the rest of this paper, we will always set $\omega$ to be zero  in non-elastic susceptibility for simplicity (see Eq.(\ref{11.1})) .

We further define the space-averaged non-elastic stress tensor and susceptibility for a single block of glass with the volume $V=L^3$, as follows, 
\begin{eqnarray}
& {} & \chi_{ijkl}(\omega) = \frac{1}{V}\int_{V} d^3xd^3x'\,\chi_{ijkl}(\vec x, \vec x'; \omega)\nonumber \\
& {} & \hat{T}_{ij} = \int_{V} d^3x\, \hat{T}_{ij}(\vec x)
\end{eqnarray}
In the rest of this paper let us use $\chi_{ijkl}$, $\chi_{ijkl}^{\rm rel}$ and $\chi_{ijkl}^{\rm res}$ to stand for $\lim_{\omega \to 0}\chi_{ijkl}(\omega)$, $\lim_{\omega\to 0}\chi_{ijkl}^{\rm rel}(\omega)$ and $\lim_{\omega\to 0}\chi_{ijkl}^{\rm res}(\omega)$, respectively. The space-averaged non-elastic susceptibility for a certain block of glass at zero-frequency limit is given as follows:

\begin{eqnarray}\label{11}
\chi_{ijkl} & = &  \chi_{ijkl}^{\rm rel}+ \chi_{ijkl}^{\rm res}\nonumber \\
\chi_{ijkl}^{\rm rel} & = & \frac{\beta}{V}\bigg(\sum_{nm}P_nP_m\langle n|\hat{T}_{ij}|n\rangle\langle m|\hat{T}_{kl}|m\rangle\nonumber \\
 & {} & \quad-\sum_n P_n\langle n|\hat{T}_{ij}|n\rangle \langle n|\hat{T}_{kl}|n\rangle\bigg)\nonumber \\
\chi_{ijkl}^{\rm res} & = & -\frac{1}{V\hbar}\sum_{n}\sum_{m\neq n}P_m\frac{\langle n|\hat{T}_{ij}|m\rangle\langle m|\hat{T}_{kl}|n\rangle}{(E_n-E_m)/\hbar+i\eta}\nonumber \\
 & {} & +
\frac{1}{V\hbar}\sum_{m}\sum_{n\neq m}P_n\frac{\langle n|\hat{T}_{ij}|m\rangle\langle m|\hat{T}_{kl}|n\rangle}{(E_n-E_m)/\hbar+i\eta}\nonumber \\
\end{eqnarray}

Finally, recall the original purpose of our paper: we want to prove the existence of positive-negative transitions in glass mechanical susceptibility when external strain passes through certain critical value (see the r.h.s. of Fig.1). The mechanical susceptibility for a single block of glass is expressed by the summation of two parts:
\begin{eqnarray}\label{5}
\chi_{ijkl}^{\rm tot} & = & \chi_{ijkl}^{\rm el}+\chi_{ijkl}
\end{eqnarray}
where for elastic susceptibility $\chi_{ijkl}^{\rm el}$ please see Eq.(\ref{7}); for non-elastic susceptibility $\chi_{ijkl}$ please see Eq.(\ref{11}). We want to investigate if there is any positive-negative transition in glass mechanical susceptibility $\chi^{\rm tot}_{ijkl}$, Eq.(\ref{5}).

\subsection{Virtual Phonon Exchange Process: Non-Elastic Stress-Stress Interaction $\hat{V}$}
The previous discussions are within the considerations of single-block glass Hamiltonian. There was not much difference between 2-level-system and generalized multiple-level-system. However, if we combine a set of $N_0^3$ single blocks together to form a ``super block", the interactions between single blocks must be taken into account. Since the non-elastic stress tensors are coupled to intrinsic strain field, if we allow virtual phonons to exchange with each other, it will generate a RKKY-type many-body interaction between single blocks. This RKKY-type interaction is the product of stress tensors of single blocks at different locations:
\begin{eqnarray}\label{12}
\hat{V}=\int d^3xd^3x'\sum_{ijkl}\Lambda_{ijkl}(\vec x-\vec x')\hat{T}_{ij}(\vec x)\hat{T}_{kl}(\vec x')
\end{eqnarray}
where the coefficient $\Lambda_{ijkl}(\vec x-\vec x')$ was first derived by J. Joffrin and A. Levelut\cite{Joffrin1976}. For detailed discussions regarding this coefficient, please see Appendix (D):

\begin{eqnarray}\label{13}
 & {} & {\Lambda}_{ijkl}(\vec x-\vec x') = \sum_{\vec k}\Lambda_{ijkl}(\vec k)e^{i\vec k\cdot (\vec x-\vec x')}\nonumber \\
 & {} & \Lambda_{ijkl}(\vec k) = \frac{\alpha}{2\rho c_t^2}\left(\frac{k_ik_jk_kk_l}{k^4}\right)\nonumber \\
 & {} & -\frac{1}{8\rho c_t^2}\left(\frac{k_jk_l\delta_{ik}+k_jk_k\delta_{il}
+k_ik_l\delta_{jk}+k_ik_k\delta_{jl}
}{k^2}\right)\nonumber \\
\end{eqnarray}
where $\alpha=1-{c_t^2}/{c_l^2}$. 
$i,j,k,l$ runs over $x,y,z$ cartesian coordinates. We name Eq.(\ref{12}) the non-elastic stress-stress interaction. In the rest of this paper for simplicity we will always use the approximation to replace $\vec x-\vec x'$ by $\vec x_{s}-\vec x_{s'}$ for the $s$-th and $s'$-th blocks, in which $\vec x_{s}$ denotes the center of the $s$-th block, and that $\int_{V^{(s)}}\hat{T}_{ij}(\vec x)d^3x=\hat{T}_{ij}^{(s)}$ is the uniform stress tensor of the $s$-th block. From this definition, the uniform stress tensor operator $\hat{T}_{ij}^{(s)}$ is a volume proportional extensive quantity. Also, from now on we use $e_{ij}^{(s)}$ to denote the strain field $e_{ij}(\vec x)$ located at the $s$-th block. By combining $N_0\times N_0\times N_0$ identical $L\times L\times L$ single blocks, we get a $N_0L\times N_0L\times N_0L$ super block. The non-elastic part of super block Hamiltonian (without the presence of external strain field) is given by the summation of single block non-elastic Hamiltonian and non-elastic stress-stress interaction:
\begin{eqnarray}\label{15}
\hat{H}_0^{\rm super}=\sum_{s=1}^{N_0^3}\hat{H}_0^{(s)}+\sum_{s\neq s'}^{N_0^3}\sum_{ijkl}\Lambda_{ijkl}^{(ss')}\hat{T}_{ij}^{(s)}\hat{T}_{kl}^{(s')}
\end{eqnarray}
From now on we apply the most important assumption of this paper: to assume that the correlation function of block uniform stress tensors $\hat{T}_{ij}^{(s)}$ are diagonal in spacial coordinates: ${\chi}^{(ss')}_{ijkl}=\frac{1}{L^3}\langle \hat{T}_{ij}^{(s)}\hat{T}_{kl}^{(s')}\rangle =\frac{1}{L^3}\langle \hat{T}_{ij}^{(s)}\hat{T}_{kl}^{(s)}\rangle\delta_{ss'}$. We assume that the uniform stress tensors lose spacial correlation for blocks at different locations $\vec x_s\neq \vec x_s'$. To further explain this assumption, we use the argument by D. C. Vural and A. J. Leggett\cite{Leggett2011}: in the absence of non-elastic stress-stress interaction $\hat{V}$, terms like $\langle n|\hat{T}_{ij}^{(s)}|m\rangle \langle m|\hat{T}_{kl}^{(s')}|n\rangle$ were originally uncorrelated. However, it is not obvious that they remain uncorrelated after the interaction $\hat{V}=\sum_{s\neq s'}^{N_0^3}\sum_{ijkl}\Lambda_{ijkl}^{(ss')}\hat{T}_{ij}^{(s)}\hat{T}_{kl}^{(s')}$ is taken into account (see super block Hamiltonian, Eq.(\ref{15})). However, we argue that on average, terms like $\langle n|\hat{T}_{ij}^{(s)}|m\rangle \langle m|\hat{T}_{kl}^{(s')}|n\rangle$ are likely to be small compared to terms like $\langle n|\hat{T}_{ij}^{(s)}|m\rangle \langle m|\hat{T}_{kl}^{(s)}|n\rangle$, because the interaction $\hat{V}$ is a highly frustrated interaction between stress tensor operators.

\subsection{Glass Non-elastic Hamiltonian with the Presence of External Strain}
In previous subsections, we used the notation $\bm{e}(\vec x)$ to stand for intrinsic strain field. In this subsection we turn on external static, uniform strain, and use the notation $\bm{e}$ to represent it. As the simplest case, we consider the external strain as $e_{xx}=e$, $e_{yy}=e_{zz}=e_{xy}=e_{yz}=e_{zx}=0$. For other kinds of external strain ${\bm e}=e_{ij}$, similar positive-negative transition behaviors of glass mechanical susceptibility will be carried out as well. We consider glass single block and super block non-elastic Hamiltonians $\hat{H}_0^{(s)}(\bm{e})$ and $\hat{H}_0^{\rm super}(\bm{e})$ with the presence of external static strain $\bm{e}$ in this subsection.

Let us combine a set of single blocks to form a super spherical glass with radius $r$. It is deformed by external strain to become an ellipsoid. The $xy$ and $xz$ plane cross sections are ellipses with eccentricity $\epsilon=\frac{\sqrt{e^2+2e}}{(1+e)}$  while the $yz$ cross section remains circular (see Fig.2 below).
\begin{figure}[hp]
\centering
\includegraphics[scale=0.35]{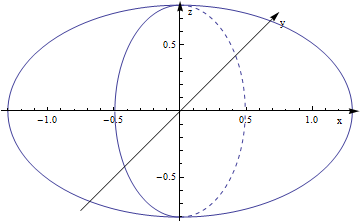}
\caption{An isotropic (spherical) glass deformed by strain $e_{xx}=e$ to become an ellipsoid.}
\end{figure}

With the presence of external static strain $\bm{e}$, the $s$-th glass single block non-elastic Hamiltonian is given by 
\begin{eqnarray}\label{16}
\hat{H}^{(s)}_0(\bm{e})=\hat{H}_0^{(s)}+e_{xx}\hat{T}_{xx}^{(s)}(\bm{e})
\end{eqnarray}
Let us define $|n(\bm{e})\rangle$ and $E_n(\bm{e})$ to be the $n$-th eigenstate and eigenvalue of single block non-elastic Hamiltonian $\hat{H}_0^{(s)}(\bm{e})$. The single block volume $V(\bm{e})=(1+e_{xx})L^3$. We define $P_n(\bm{e})=e^{-\beta E_n(\bm{e})}/\mathcal{Z}(\bm{e})$ to be the probability function of $n$-th eigenstate of $\hat{H}_0^{(s)}(\bm{e})$. The glass single block zero-frequency non-elastic susceptibility $\chi_{ijkl}(\bm{e})$ is given by 
\widetext
\begin{eqnarray}\label{17}
\chi_{ijkl}(\bm{e}) & = &  \chi_{ijkl}^{\rm rel}(\bm{e})+ \chi_{ijkl}^{\rm res}(\bm{e})\nonumber \\
\chi_{ijkl}^{\rm rel}(\bm{e}) & = & \frac{\beta}{V(\bm{e})}\bigg(\sum_{nm}P_n(\bm{e})P_m(\bm{e})\langle n(\bm{e})|\hat{T}_{ij}(\bm{e})|n(\bm{e})\rangle\langle m(\bm{e})|\hat{T}_{kl}(\bm{e})|m(\bm{e})\rangle\nonumber \\
 & {} & \qquad\,\,-\sum_n P_n(\bm{e})\langle n(\bm{e})|\hat{T}_{ij}(\bm{e})|n(\bm{e})\rangle \langle n(\bm{e})|\hat{T}_{kl}(\bm{e})|n(\bm{e})\rangle\bigg)\nonumber \\
\chi_{ijkl}^{\rm res}(\bm{e}) & = & \frac{1}{V(\bm{e})}\sum_{n}\sum_{m\neq n}P_n(\bm{e})\frac{\langle n(\bm{e})|\hat{T}_{ij}(\bm{e})|m(\bm{e})\rangle\langle m(\bm{e})|\hat{T}_{kl}(\bm{e})|n(\bm{e})\rangle}{E_n(\bm{e})-E_m(\bm{e})+i\eta}\nonumber \\
 & - & \frac{1}{V(\bm{e})}\sum_{m}\sum_{n\neq m}P_m(\bm{e})\frac{\langle n(\bm{e})|\hat{T}_{ij}(\bm{e})|m(\bm{e})\rangle\langle m(\bm{e})|\hat{T}_{kl}(\bm{e})|n(\bm{e})\rangle}{E_n(\bm{e})-E_m(\bm{e})+i\eta}
\end{eqnarray}
For details of obtaining the results, Eq.(\ref{16}, \ref{17}), please see Appendix (A). 
\endwidetext

On the other hand, the super block glass non-elastic Hamiltonian with the presence of external strain $\bm{e}$ is given by 
\begin{eqnarray}\label{18}
\hat{H}_0^{\rm super}(\bm{e}) & = & \sum_{s}\bigg(\hat{H}_0^{(s)}+\sum_{ij}e_{xx}\hat{T}_{xx}^{(s)}(\bm{e})\bigg)\nonumber \\
 & + & \sum_{s\neq s'}\sum_{ijkl}\Lambda_{ijkl}^{(ss')}(\bm{e})\hat{T}_{ij}^{(s)}(\bm{e})\hat{T}_{kl}^{(s')}(\bm{e})
\end{eqnarray}
We denote $|n^*(\bm{e})\rangle$ and $E_n^*(\bm{e})$ to be the $n^*$-th eigenstate and eigenvalue of $\hat{H}_0^{\rm super}(\bm{e})$, which will be useful in later discussions.

According to the definition of non-elastic stress tensor, Eq.(\ref{9}), the super block non-elastic stress tensor is defined as the derivative of $\hat{H}^{\rm super}(\bm{e})$ with respect to infinitesimal intrinsic strain: $\hat{T}_{ij}^{\rm super}(\bm{e})={\delta  \hat{H}^{\rm super}(\bm{e})}/{\delta e_{ij}}$. Because of the strain field dependence of $\Lambda_{ijkl}^{(ss')}(\bm{e})$, an extra term proportional to ${\delta \Lambda_{abcd}^{(ss')}(\bm{e})}/{\delta e_{ij}}$ appears in super block stress tensor:
\begin{eqnarray}\label{19}
\hat{T}_{ij}^{\rm super}(\bm{e}) & = & \sum_s\hat{T}_{ij}^{(s)}(\bm{e}) +\sum_{ss'}\sum_{abcd}\frac{\delta \Lambda_{abcd}^{(ss')}(\bm{e})}{\delta e_{ij}}\hat{T}_{ab}^{(s)}(\bm{e})\hat{T}_{cd}^{(s')}(\bm{e})\nonumber \\
\end{eqnarray}

Smilar with super block non-elastic stress tensor, the super block non-elastic susceptibility also receives an extra term. The super block non-elastic susceptibility is given as follows 
\widetext
\begin{eqnarray}\label{20}
\chi_{ijkl}^{\rm super}(\bm{e})
 & = & \frac{\beta}{N_0^3V(\bm{e})}\bigg(\sum_{n^*m^*}P_n^*(\bm{e})P_m^*(\bm{e})\langle n^*(\bm{e})|\sum_s\hat{T}^{(s)}_{ij}(\bm{e})|n^*(\bm{e})\rangle\langle m^*(\bm{e})|\sum_{s'}\hat{T}^{(s')}_{kl}(\bm{e})|m^*(\bm{e})\rangle\nonumber \\
 & {} & \qquad\qquad -\sum_{n} P_n^*(\bm{e})\langle n^*(\bm{e})|\sum_{s}\hat{T}^{(s)}_{ij}(\bm{e})|n^*(\bm{e})\rangle \langle n^*(\bm{e})|\sum_{s'}\hat{T}^{(s')}_{kl}(\bm{e})|n^*(\bm{e})\rangle\bigg)\nonumber \\
 & + & \frac{1}{N_0^3V(\bm{e})}\bigg(\sum_{n^*}\sum_{m^*\neq n^*}P_n^*(\bm{e})\frac{\langle n^*(\bm{e})|\sum_s\hat{T}^{(s)}_{ij}(\bm{e})|m^*(\bm{e})\rangle\langle m^*(\bm{e})|\sum_{s'}\hat{T}^{(s')}_{kl}(\bm{e})|n^*(\bm{e})\rangle}{E_n^*(\bm{e})-E^*_m(\bm{e})+i\eta}\nonumber \\
 & {} & \qquad\qquad -\sum_{m^*}\sum_{n^*\neq m^*}P_m^*(\bm{e})\frac{\langle n^*(\bm{e})|\sum_s\hat{T}^{(s)}_{ij}(\bm{e})|m^*(\bm{e})\rangle\langle m^*(\bm{e})|\sum_{s'}\hat{T}^{(s')}_{kl}(\bm{e})|n^*(\bm{e})\rangle}{E_n^*(\bm{e})-E^*_m(\bm{e})+i\eta}\bigg)\nonumber \\
 & + & \frac{1}{N_0^3V(\bm{e})}\sum_{abcd}\sum_{ss'}\left<\frac{\delta^2\Lambda_{abcd}^{(ss')}(\bm{e})}{\delta e_{ij}\delta e_{kl}}\hat{T}_{ab}^{(s)}(\bm{e})\hat{T}_{cd}^{(s')}(\bm{e})\right>
\end{eqnarray}
\endwidetext
In the above result, $P_n^*(\bm{e})=e^{-\beta E_n^*(\bm{e})}/\mathcal{Z}^*(\bm{e})$ is the probability function of the $n^*$-th eigenstate of $\hat{H}_0^{\rm super}(\bm{e})$. The first and second lines stand for super block zero-frequency relaxation and resonance susceptibilities, while the third line is the extra term which comes from the second term of Eq.(\ref{19}). For details of discussions on super block non-elastic Hamiltonian and susceptibility with the presence of external strain, please see Appendix (B). In the next section we will prove that the last term in Eq.(\ref{20}) in real space renormalization irrelevant.

\section{Real Space Renormalization Procedure of Glass Non-Elastic Susceptibility}
Let us combine $N_0^3$ glass single blocks with the dimension $L\times L\times L$ to form a super block with the dimension $N_0L\times N_0L\times N_0L$. Because the super block is $N_0^3$ times the volume of single block, repeating such process from microscopic length scale will eventually carry out the glass Hamiltonian, non-elastic stress tensor and susceptibility at experimental length scale. Our main purpose of this section is to set up the recursion relation (i.e., real space renormalization equation) between single block and super block non-elastic susceptibilities.

According to the argument by D. C. Vural and A. J. Leggett\cite{Leggett2011}, the suggested starting microscopic length scale of renormalization procedure is, for example, $L_1\sim 50\AA$. Since the final result only logarithmically depends on this choice, it will not be sensitive. The chacteristic thermal phonon wavelength must be no less than the starting length scale $L_1$, which means the effective temperature of our model is no greater than $60$K. However, at least to the author's knowledge, all of the experiments on glass avalanche are taken under room temperatures or glass transition temperatures\cite{Argon2005, Gauthier2004, Dahmen2014, Fineberg1991, Fineberg1992} ($T\sim 300$K). We hope more experiments on mechanical properties of glass could be taken at low-temperatures below 60K to further test the validity of our theory.

In the $n$-th step renormalization, we combine $N_0^3$ single blocks with the side $L_n$ to form the $n$-th step super block glass with the side $N_0L_n$.  The single block in the $(n+1)$-th step is actually the super block in the $n$-th step, i.e., $L_{n+1}=N_0L_n$. Before we turn on non-elastic stress-stress interaction $\hat{V}(\bm{e})$, such a group of non-interacting single blocks have the Hamiltonian $\sum_{s=1}^{N_0^3}\left(\hat{H}^{(s)}_0+e_{xx}\hat{T}_{xx}^{(s)}(\bm{e})\right)$, eigenstates $|n(\bm{e})\rangle = \prod_{s=1}^{N_0^3}|n^{(s)}(\bm{e})\rangle$ and eigenvalues $E_n(\bm{e})=\sum_{s=1}^{N_0^3}E_n^{(s)}(\bm{e})$, where $|n^{(s)}(\bm{e})\rangle$ and $E_n^{(s)}(\bm{e})$ are the $n^{(s)}$-th eigenstate and eigenvalue of the $s$-th single block non-elastic Hamiltonian $\hat{H}_0^{(s)}(\bm{e})=\hat{H}_0^{(s)}+e_{xx}\hat{T}_{xx}^{(s)}(\bm{e})$. We combine single blocks into a super block and turn on interaction $\hat{V}(\bm{e})=\sum_{s\neq s'}^{N_0^3}\Lambda_{ijkl}^{(ss')}(\bm{e})\hat{T}^{(s)}_{ij}(\bm{e})\hat{T}^{(s')}_{kl}(\bm{e})$. We assume interaction $\hat{V}(\bm{e})$ is relatively weak compared to $\hat{H}_0(\bm{e})=\sum_{s=1}^{N_0^3}\left(\hat{H}^{(s)}_0+e_{xx}\hat{T}_{xx}^{(s)}(\bm{e})\right)$, so that the interaction can be treated as a perturbation. In fact, non-elastic stress-stress interaction is $\hat{V}(\bm{e})\propto 1/r^3$. At small length scales we know it always dominates glass Hamiltonian. However, if the non-elastic susceptibility decreases logarithmically as the increase of length scale (which will be proved in Eq.(\ref{3.23.4})), then that means the non-elastic stress-stress interaction $\hat{V}(\bm{e})$ can be treated as a perturbation at the late stages. The assumption that $\hat{V}(\bm{e})$ can be treated as a perturbation is qualitatively correct. In the last section we define the eigenstates and eigenvalues of super block non-elastic Hamiltonian $\hat{H}_0^{\rm super}(\bm{e})$ to be $|n^*(\bm{e})\rangle$ and $E_n^*(\bm{e})$. Their relations with $|n(\bm{e})\rangle$ and $E_n(\bm{e})$ are given as 

\begin{eqnarray}\label{21}
|n^*(\bm{e})\rangle & = & |n(\bm{e})\rangle +\sum_{p\neq n}\frac{\langle p(\bm{e})|\hat{V}(\bm{e})|n(\bm{e})\rangle}{E_n(\bm{e})-E_p(\bm{e})}|p(\bm{e})\rangle+\mathcal{O}(V^2)\nonumber \\
E_n^*(\bm{e}) & = & E_n(\bm{e})+\langle n(\bm{e})|\hat{V}(\bm{e})|n(\bm{e})\rangle +\mathcal{O}(V^2)
\end{eqnarray}
With the relations in Eq.(\ref{21}) one can expand super block non-elastic susceptibility Eq.(\ref{20}) in orders of $\hat{V}(\bm{e})$. Up to the first order in $\hat{V}(\bm{e})$ we can write super block non-elastic susceptibility ${\chi}^{\rm super}_{ijkl}(\bm{e})$ expansions in terms of single block non-elastic susceptibilities ${\chi}_{ijkl}(\bm{e})$. The recursion relation for single block and super block non-elastic susceptibilities are given as follows:

\widetext
\begin{eqnarray}\label{23}
\chi_{ijkl}^{\rm super}(\bm{e}) & = & \chi_{ijkl}^{\rm super\,rel}(\bm{e})+\chi_{ijkl}^{\rm super\, res}(\bm{e})\nonumber \\
 & = & 
\chi_{ijkl}^{\rm rel}(\bm{e})
 -\frac{V_n(\bm{e})}{N_0^3}\left[-\sum_{mnpq}\sum_{ss'}\Lambda_{mnpq}^{(ss')}(\bm{e})\right]\left(\chi_{ijmn}^{\rm rel}(\bm{e})\chi_{pqkl}^{\rm rel}(\bm{e})+\chi_{ijmn}^{\rm rel}(\bm{e})\chi_{pqkl}^{\rm res}(\bm{e})+\chi_{ijmn}^{\rm res}(\bm{e})\chi_{pqkl}^{\rm rel}(\bm{e})\right)\nonumber \\
 & + & \chi_{ijkl}^{\rm res}(\bm{e}) -\frac{V_n(\bm{e})}{N_0^3}\left[-\sum_{mnpq}\sum_{ss'}\Lambda_{mnpq}^{(ss')}(\bm{e})\right] \chi_{ijmn}^{\rm res}(\bm{e})\chi_{pqkl}^{\rm res}(\bm{e})\nonumber \\
 & + & \frac{1}{N_0^3V_n(\bm{e})}\sum_{mnpq}\sum_{ss'}\left<\frac{\delta^2\Lambda_{mnpq}^{(ss')}(\bm{e})}{\delta e_{ij}\delta e_{kl}}\hat{T}_{mn}^{(s)}(\bm{e})\hat{T}_{pq}^{(s')}(\bm{e})\right>
\end{eqnarray}
\endwidetext
For details of calculations of obtaining Eq.(\ref{23}), please see Appendix (C). The last term of Eq.(\ref{23}) is renormalization irrelevant. Compared to other terms in Eq.(\ref{23}) the last term decreases  ($\propto L^{-3}$) as the increase of length scale $L$. To prove this result let us provide a qualitative analysis: denote $\Lambda_{ijkl}^{(ss')}(\bm{e})=-\tilde{\Lambda}_{ijkl}(\bm{e})/8\pi\rho c_t^2x_{ss'}^3$ where $x_{ss'}=|\vec x_s-\vec x_s'|$ and $\tilde{\Lambda}_{ijkl}(\bm{e})$ is a dimensionless number of order 1 (see Eq.(\ref{13})). By applying linear response theory on the last term of Eq.(\ref{23}) with respect to infinitesimal external perturbation $\sum_{ij}\sum_se_{ij}^{(s)}(t)\hat{T}_{ij}^{(s)}$, the last term of Eq.(\ref{23}) is given by

\begin{eqnarray}\label{23.1}
\sum_{mnpq}\int d\Omega\,{\rm Im}\,{\chi}_{ijmn}^{\rm res}(\Omega)\left(\sum_{ss'}\frac{\hbar L_n^3{\lambda}_{mnpq}}{ 8\pi\rho^2 c_t^4x_{ss'}^6}\right)\,{\rm Im}\,{\chi}^{\rm res}_{pqkl}(-\Omega)\nonumber \\
\end{eqnarray}
where $\lambda_{mnpq}(\bm{e})$ is the second order derivative of $\tilde{\Lambda}_{mnpq}(\bm{e})$ with respect to strain. It is also a dimensionless number of order 1. According to the argument by R. O. Pohl and his group and D. C. Vural and A. J. Leggett\cite{Pohl2002, Leggett2011}, we assume that the ``reduced imaginary part of non-elastic resonance susceptibility", ${\rm Im}\,\tilde{\chi}_{l,t}^{\rm res}(\omega, T)=\left(1-e^{-\beta\hbar\omega}\right)^{-1}\,{\rm Im}\,\chi_{l,t}^{\rm res}(\omega, T)$ is approximately a constant of frequency and temperature up to $\omega_c\sim 10^{15}$Hz and around the temperature of order 1K. Since the imaginary part of resonance susceptibility is always smaller than it's reduced version: ${\rm Im\,}{\chi}^{\rm res}_{ijkl}(\omega)<{\rm Im\,}\tilde{\chi}^{\rm res}_{ijkl}(\omega)$, integrating over $\Omega$ gives the upper limit of Eq.(\ref{23.1}): $-{\rm C}{\hbar\omega_c}\,\left({\rm Im}\,\tilde{\chi}_t^{\rm res}\right)^2/{ \rho^2 c_t^4L_n^3}$, where ${\rm C}$ is also a dimensionless constant of order 1. If we require that there is a critical length scale $L_c$, below which the last term of Eq.(\ref{23}) is comparable to the other terms, the order of magnitude for $L_c$ is 
\begin{eqnarray}\label{23.2}
L_c< \left(\frac{\hbar\omega_c}{\rho c_{l,t}^2}\right)^{\frac{1}{3}}\approx 4.6\AA< L_1=50\AA
\end{eqnarray}
which means the upper limit of $L_c$ is even smaller than the starting effective length scale of renormalization technique. Throughout the entire renormalization procedure the last term in Eq.(\ref{23}) is always negligible. In the following discussions we drop the last term in Eq.(\ref{23}).

With the above simplifications one can rewrite the non-elastic susceptibility renormalization equation as follows, 
\begin{eqnarray}\label{3.23.1}
 & {} & \chi_{ijkl}^{\rm super}(\bm{e}) 
= \nonumber \\
 & {} & \chi_{ijkl}(\bm{e}) 
 -\frac{V_n(\bm{e})}{N_0^3}\left[-\sum_{mnpq}\sum_{ss'}\Lambda_{mnpq}^{(ss')}(\bm{e})\right]\chi_{ijmn}(\bm{e}) \chi_{pqkl}(\bm{e}) \nonumber \\
\end{eqnarray}
where the zero-frequency non-elastic susceptibilities are given by $\chi_{ijkl}(\bm{e}) =\chi_{ijkl}^{\rm rel}(\bm{e}) +\chi_{ijkl}^{\rm res}(\bm{e}) $, and $\chi_{ijkl}^{\rm super}(\bm{e}) =\chi_{ijkl}^{\rm super\,rel}(\bm{e}) +\chi_{ijkl}^{\rm super\,res}(\bm{e}) $.

The renormalization equation for non-elastic susceptibility can be further simplified with the following three steps. First of all, we define a 4-indice quantity $M_{mnpq}$, given by 
\begin{eqnarray}\label{3.23.3}
M_{mnpq}(\bm{e}) =\frac{V_n(\bm{e})}{N_0^3}\left[-\sum_{ss'}\Lambda_{mnpq}^{(ss')}(\bm{e})\right]
\end{eqnarray}
So the non-elastic susceptibility renormalization relation is simplified as 
\begin{eqnarray}\label{3.23.2}
\chi_{ijkl}^{\rm super} (\bm{e}) 
 & = & 
\chi_{ijkl}(\bm{e}) 
 -\sum_{mnpq}\chi_{ijmn}(\bm{e}) M_{mnpq}(\bm{e}) \chi_{pqkl}(\bm{e}) \nonumber \\
\end{eqnarray}

Second, let us denote the 2-fold indices in Eq.(\ref{3.23.2}), $(ij), (kl), (mn), (pq)$  to be $(ij)\to A$, $(kl)\to B$, $(mn)\to C$, $(pq)\to D$. With this simplification, we rewrite 4-indice quantities $\chi_{ijkl}$ and $M_{mnpq}$ into 2-indice matrix forms: $\bm{\chi}(\bm{e})={\chi}_{AB}(\bm{e})$ and $\bm{M}(\bm{e})={M}_{CD}(\bm{e})$. They are $6\times 6$ matrices, for example, $M_{CD}$ has the indices $C \,({\rm or}\, D)=(xx), (xy), (xz), (yy), (yz), (zz)$. Third, let us define the ``change of non-elastic susceptibility", $\delta\bm{\chi}(\bm{e})=\bm{\chi}^{\rm super}(\bm{e})-\bm{\chi}(\bm{e})$. We integrate over the length scale to calculate Eq.(\ref{3.23.2}). The macroscopic length scale non-elastic susceptibility is therefore given as follows:
\begin{eqnarray}\label{3.23.4}
\bm{\chi}^{-1}(R)=\bm{M}(\bm{e})\log_{N_0}\left(\frac{R}{L_1}\right)+\bm{\chi}^{-1}(L_1)
\end{eqnarray}
where the experimental length scale $R$ is the actual size of glass sample. We have no idea about the value of the starting microscopic non-elastic susceptibility $\bm{\chi}(L_1)$. However, we would like to argue, that the term $\bm{M}(\bm{e})\log_{N_0}\left({R}/{L_1}\right)$ is much greater than $\bm{\chi}^{-1}(L_1)$ becasue of the factor $\log_{N_0}\left({R}/{L_1}\right)$. In the rest of this paper, we neglect the starting microscopic length scale non-elastoc susceptibility $\bm{\chi}(L_1)$ in Eq.(\ref{3.23.4}).


\section{The Positive-Negative Transitions of Non-elastic Susceptibility}
We want to prove the existence of positive-negative transitions in glass mechanical susceptibility with the increase of external static strain $\bm{e}$. Since the positive elastic susceptibility $\chi_{ijkl}^{\rm el}=(\rho c_l^2-2\rho c_t^2)\delta_{ij}\delta_{kl}+\rho c_t^2(\delta_{ik}\delta_{jl}+\delta_{il}\delta_{jk})$ is a constant, our hope is to find positive-negative transitions in non-elastic susceptibility, Eq.(\ref{11}). This is the main purpose of this section.

The spherical glass is deformed by external static strain $\bm{e}=e_{xx}$ to become an ellipsoid with the eccentricity $\epsilon=\sqrt{({e^2+2e})/({1+e})}$ in $xy, xz$ plane cross sections. In Eq.(\ref{3.23.3}) let us take continuum limit and change the variables of spacial integrals $\vec r_s+\vec r_s'=\vec R$ and $\vec r_s-\vec r_s'=\vec r$ to calculate the 4-indice quantity $M_{mnpq}(\bm{e})$:
\begin{eqnarray}\label{26}
M_{mnpq}(\bm{e})=-\int_{V(\bm{e})} d^3r\,{\Lambda}_{mnpq}(\vec x_s-\vec x_s')
\end{eqnarray}
where the integral region $V(\bm{e})$ is the space of ellipsoid glass. After integrating over the ellipsoid region, the quantity $\bm{M}(\bm{e})\log_{N_0}(R/L_1)$ which appears in Eq.(\ref{3.23.4}) is then given by the following form
\widetext
\begin{eqnarray}\label{27}
 & {} & \bm{M}(\bm{e})\log_{N_0}\left(\frac{R}{L_1}\right)=\frac{1}{2\rho c_t^2}\ln\left(\frac{R}{L_1}\right)\nonumber \\
 & {} & \left(\begin{array}{cccccc}
A+\frac{4\pi}{3}\left(1-\frac{1}{5}\alpha\right) & 0 & 0 & B-\frac{8\pi}{90}\alpha & 0 & B-\frac{8\pi}{90}\alpha\\
 0 & C+\frac{2\pi}{3}\left(1-\frac{4}{30}\alpha\right) & 0 & 0 & 0 & 0\\
 0 & 0 & C+\frac{2\pi}{3}\left(1-\frac{4}{30}\alpha\right) & 0 & 0 & 0\\
B-\frac{8\pi}{90}\alpha & 0 & 0 & D+\frac{4\pi}{3}\left(1-\frac{1}{5}\alpha\right) & 0 & E-\frac{8\pi}{90}\alpha\\
 0 & 0 & 0 & 0 & F+\frac{2\pi}{3}\left(1-\frac{4}{30}\alpha\right) & 0\\
B-\frac{8\pi}{90}\alpha & 0 & 0 & E-\frac{8\pi}{90}\alpha & 0 & D+\frac{4\pi}{3}\left(1-\frac{1}{5}\alpha\right)\\
\end{array}\right)\nonumber \\
\end{eqnarray}
\endwidetext
To obtain the above matrix form of $\bm{M}(\bm{e})$, first we write the 4-indice quantity $M_{mnpq}(\bm{e})$ as $M_{(mn)(pq)}(\bm{e})$. Then we calculate the entire 36 matrix elements of it to obtain the above result, Eq.(\ref{27}). $A$, $B$, $C$, $D$, $E$ and $F$ are given as follows, 
\begin{eqnarray}\label{28}
A & = & 1-3\overline{n_x^2}+\frac{1}{2}\alpha\left(-3+18\overline{ n_x^2}-15\overline{ n_x^4}\right)\nonumber \\
B & = & \frac{1}{2}\alpha\left[-1-15\overline{ n_x^2n_y^2}+3\left(\overline{ n_x^2}+\overline{ n_y^2}\right)\right]\nonumber \\
C & = & \frac{1}{4}\left(2-3\overline{ n_x^2}-3\overline{ n_y^2}\right)\nonumber \\
 & {} & +\frac{1}{2}\alpha\left[-1-15\overline{ n_x^2n_y^2}+3\left(\overline{ n_x^2}+\overline{ n_y^2}\right)\right]\nonumber \\
D & = & 1-3\overline{ n_y^2}+\frac{1}{2}\alpha\left(-3+18\overline{ n_y^2}-15\overline{ n_y^4}\right)\nonumber \\
E & = & \frac{1}{2}\alpha\left(-1-15\overline{ n_y^2n_z^2}+6\overline{ n_y^2}\right)\nonumber \\
F & = & \frac{1}{2}\left(1-3\overline{ n_y^2}\right)+\frac{1}{2}\alpha\left(-1-15\overline{ n_y^2n_z^2}+6\overline{ n_y^2}\right)
\end{eqnarray}
In the above result we have applied rotational invariance of the integral region $V(\bm{e})$ around $x$-axis, and the experimentally measurable quantity $\alpha$ is $\alpha=1-c_t^2/c_l^2$. The definition of average values $\overline{ n_{x,y}^2}$, $\overline{ n_{x,y}^4}$, $\overline{ n_x^2n_y^2}$, $\overline{ n_y^2n_z^2}$ are given as follows: for an arbitrary function $f(\vec r)$, the average value $\overline{ f(\vec r)}$ is defined as  
\begin{eqnarray}\label{26}
\overline{ f(\vec r)}=\frac{\int_{V(\bm{e})} d^3r \, f(\vec r)/r^3}{\int_{V(\bm{e})} d^3r \, 1/r^3}
\end{eqnarray}
Integrate over the ellipsoid region $V(\bm{e})$, the unit vector average values are displayed as follows,

\begin{eqnarray}\label{29}
 & {} & \overline{ n_x^2}  = \frac{\epsilon\sqrt{1-\epsilon^2}(-1+2\epsilon^2)+\arcsin\epsilon}
{4\epsilon^2\left(\epsilon\sqrt{1-\epsilon^2}+\arcsin\epsilon\right)}\nonumber \\
 & {} & \overline{ n_x^4}  = \frac{\epsilon\sqrt{1-\epsilon^2}(-3-2\epsilon^2+8\epsilon^4)+3\arcsin\epsilon}
{24\epsilon^4\left(\epsilon\sqrt{1-\epsilon^2}+\arcsin\epsilon\right)}\nonumber \\
 & {} & \overline{ n_y^2n_z^2} =
\frac{\epsilon\sqrt{1-\epsilon^2}(-3+10\epsilon^2+8\epsilon^4)}{192\epsilon^4\left(\epsilon\sqrt{1-\epsilon^2}+\arcsin\epsilon\right)}\nonumber \\
 & {} & \quad\quad\quad\,\,\,+\frac{3(1-4\epsilon^2+8\epsilon^4)\arcsin\epsilon}{192\epsilon^4\left(\epsilon\sqrt{1-\epsilon^2}+\arcsin\epsilon\right)}
\end{eqnarray}
where $0\le\epsilon\le 1$ is the eccentricity of the $xy$ and $xz$ cross sections of ellipsoid glass (see Fig.2).

Eq.(\ref{27}) is a $6\times 6$ matrix, so it has 6 eigenvalues and eigenvectors. We want to figure out which of these eigenvalues show positive-negative transitions. First of all we list a series of variable changes:
\begin{eqnarray}\label{30}
 & {} & A' = A+\frac{4\pi}{3}\left(1-\frac{1}{5}\alpha\right)\quad\,\,\,\, B' = B-\frac{8\pi}{90}\alpha\nonumber \\
 & {} & C' = C+\frac{2\pi}{3}\left(1-\frac{4}{30}\alpha\right)\quad D' = D+\frac{4\pi}{3}\left(1-\frac{1}{5}\alpha\right)\nonumber \\
 & {} & E' = E-\frac{8\pi}{90}\qquad \qquad\qquad\,\,\, F' = F+\frac{2\pi}{3}\left(1-\frac{4}{30}\alpha\right)\nonumber \\
 & {} & \Delta' = 8B'^2+(A'-D'-E')^2
\end{eqnarray}

Among the 6 eigenvalues of $\bm{M}(\bm{e})\log_{N_0}(\frac{R}{L_1})=\bm{\chi}^{-1}(R)$, 3 of them keep positive, while other 3 show positive-negative transitions we demand. The eigenvalues and corresponding eigenvectors of $\bm{\chi}^{-1}(R)$ are listed as follows:
\begin{eqnarray}\label{31}
\begin{array}{|c|c|}
\hline
 {\rm eigenvalue} & {\rm eigenvector}\\
\hline
C' & (0,0,1,0,0,0)\\
C' & (0,1,0,0,0,0)\\
({A'+D'+E'+\sqrt{\Delta'}})/2 & \left(\frac{A'-D'-E'+\sqrt{\Delta'}}{2B'},0,0,1,0,1\right)\\
({A'+D'+E'-\sqrt{\Delta'}})/2 & \left(\frac{A'-D'-E'-\sqrt{\Delta'}}{2B'},0,0,1,0,1\right)\\
D'-E' & (0,0,0,-1,0,1)\\
F' & (0,0,0,0,1,0)\\
\hline
\end{array}
\nonumber \\
\end{eqnarray}
The first three eigenvalues $C', C'$ and $(A'+D'+E'+\sqrt{\Delta'})/2$ stay positive for arbitrary positive $\chi_{l,t}(L_1)$ and $\epsilon$; other three eigenvalues, $(A'+D'+E'-\sqrt{\Delta'})/2$, $D'-E'$ and $F'$ decrease from positive to negative with the increase of eccentricity.

We choose the typical value of $\alpha=1-c_t^2/c_l^2=0.7$ and $R=1$m, so $\ln(R/L_1)\approx 20$. 
The 6 eigenvalues of non-elastic susceptibility $\bm{\chi}(R)$ are $C'^{-1}$, $C'^{-1}$, ${2}/({A'+D'+E'+\sqrt{\Delta'}})$, ${2}/({A'+D'+E'-\sqrt{\Delta'}}) $, $\left(D'-E'\right)^{-1}$ and $F'^{-1}$. The first, second and third eigenvalues $C'^{-1}$, $C'^{-1}$ and ${2}/({A'+D'+E'+\sqrt{\Delta'}})$ stay positive for eccentricities vary from 0 to 1. The plots of eigenvalues versus eccentricities are displayed in Figs.3 and 4. As the external static deformation $e_{xx}=e$ increases, glass is hardening against the external strains in the directions of $e_{xy}$, $e_{xz}$ and $\frac{A'-D'-E'+\sqrt{\Delta'}}{2B'}e_{xx}+e_{yy}+e_{zz}$. The coefficient $\frac{A'-D'-E'+\sqrt{\Delta'}}{2B'}$ stays negative for arbitrary eccentricity varies from 0 to 1. We plot the coefficient $\frac{A'-D'-E'+\sqrt{\Delta'}}{2B'}$ in Fig.5 below.

\begin{figure}[hp]
\centering
\includegraphics[scale=0.4]{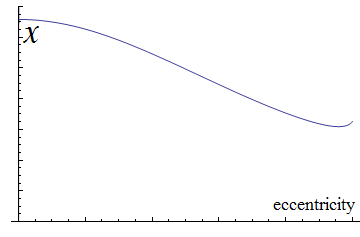}
\caption{The first and second eigenvalues of non-elastic susceptibility, $C'^{-1}$ as the function of eccentricity. It stays positive.}
\end{figure}

\begin{figure}[hp]
\centering
\includegraphics[scale=0.4]{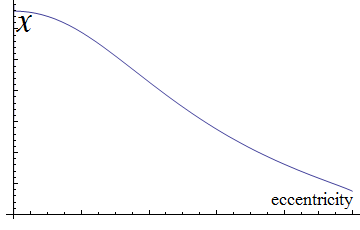}
\caption{The third eigenvalue $2/(A'+D'+E'+\sqrt{\Delta'})$ of non-elastic susceptibility as the function of eccentricity. It stays positive.}
\end{figure}

\begin{figure}[hp]
\centering
\includegraphics[scale=0.4]{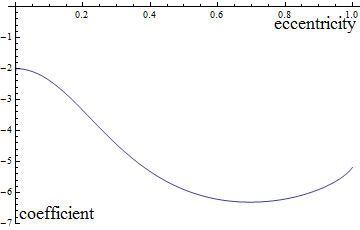}
\caption{The \textbf{coefficient} in the third eigenvector, $\frac{A'-D'-E'+\sqrt{\Delta'}}{2B'}$ as the function of eccentricity. It stays negative.}
\end{figure}

On the other hand, the fourth, fifth and sixth eigenvalues of non-elastic susceptibility $\bm{\chi}(R)$, $2/({A'+D'+E'-\sqrt{\Delta}})$, $\left(D'-E'\right)^{-1}$ and $F'^{-1}$ present positive-negative transitions at certain critical eccentricities. We plot them as the function of eccentricity in Figs. 6, 7 and 8 as follows. The coefficient of the fourth eigenvector, $\frac{A'-D'-E'-\sqrt{\Delta}}{2B'}$ is always positive. We show this plot in Fig.9. Fig.6-8 indicate when external static deformations $e_{xx}=e$ exceed certain critical values, glass is fragile against the external strain fields in the directions of $\frac{A'-D'-E'-\sqrt{\Delta}}{2B'}e_{xx}+e_{yy}+e_{zz}$, $-e_{yy}+e_{zz}$ and $e_{yz}$.

\begin{figure}[hp]
\centering
\includegraphics[scale=0.45]{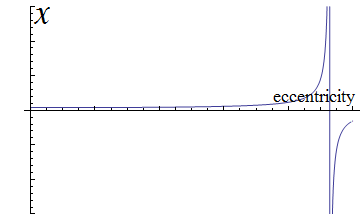}
\caption{The fourth eigenvalue of non-elastic susceptibility, $2/({A'+D'+E'-\sqrt{\Delta}})$ as the function of eccentricity. It presents a sudden positive-infinite to negative-infinite transition at the critical eccentricity $\epsilon_{\rm crit}^{(1)}=1.70$ (with the input of parameters we mentioned before). }
\end{figure}

\begin{figure}[hp]
\centering
\includegraphics[scale=0.45]{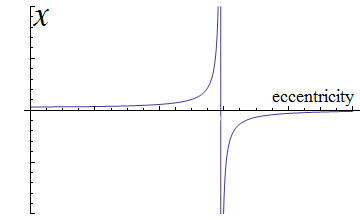}
\caption{The fifth eigenvalue of non-elastic suscptibility, $\left(D'-E'\right)^{-1}$ as the function of eccentricity. It presents a sudden positive-infinite to negative-infinite transition at the critical eccentricity $\epsilon_{\rm crit}^{(2)}=0.239$. }
\end{figure}

\begin{figure}[hp]
\centering
\includegraphics[scale=0.45]{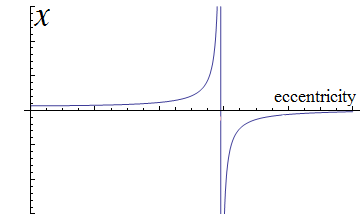}
\caption{The sixth eigenvalue $F'^{-1}$ as the function of eccentricity. It presents a positive-negative transition at the critical eccentricity $\epsilon_{\rm crit}^{(3)}=\epsilon_{\rm crit}^{(2)}=0.239$. }
\end{figure}

\begin{figure}[hp]
\centering
\includegraphics[scale=0.45]{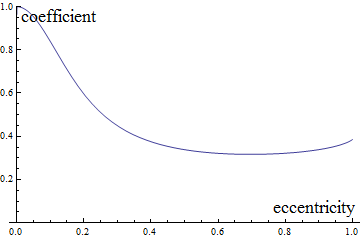}
\caption{The \textbf{coefficient} in the sixth eigenvector, $\frac{A'-D'-E'-\sqrt{\Delta}}{2B'}$ as the function of eccentricity. It stays positive with the value varying from $0.4$ to $1$.}
\end{figure}

Let us discuss the fourth, fifth and sixth eigenvalues of non-elastic susceptibility in details. First of all, the eigensvector which corresponds to the fourth eigenvalue is $\left(\frac{A'-D'-E'-\sqrt{\Delta}}{2B'},0,0,1,0,1\right)$. From Fig.9, the coefficients of $e_{xx}$, $e_{yy}$ and $e_{zz}$ have the same signs. When the external static strain $e_{xx}$ exceeds critical value $e_{\rm crit}^{(1)}=1.70$, glass is fragile against an additional expansion or contraction. Second, the eigenvector which corresponds to the fifth eigenvalue is $(0,0,0,-1,0,1)$. When the external static strain exceeds critical value $e_{\rm crit}^{(2)}=0.239$, glass is fragile against an additional strain $\pm(e_{yy}-e_{zz})$. Third, the eigenvector which corresponds to the sixth eigenvalue is $e_{yz}$. When the external static strain exceeds critical value $e_{\rm crit}^{(3)}=e_{\rm crit}^{(2)}=0.239$, glass is fragile against an additional shear deformation in $yz$ plane. Since the glass is symmstric around the $x$-axis, it is obvious that the critical value should be the same for external strain fields $\pm(e_{yy}-e_{zz})$ and $e_{yz}$. However, compared to $e_{\rm crit}^{(2), (3)}=0.239$, the first critical strain $e_{\rm crit}^{(1)}=1.70$ is too big to be observed. Before reaching such great value, glass has already reached the other two critical strains to break. From the above discussions, we would like to predict, that the amorphous materials are more likely to crack for shear deformations rather than dilation deformations.

\widetext

\begin{figure}[hp]
\centering
\includegraphics[scale=0.35]{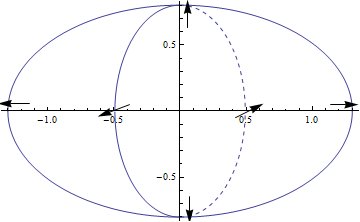}
\includegraphics[scale=0.35]{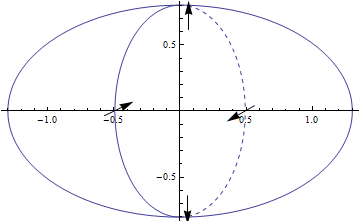}
\includegraphics[scale=0.35]{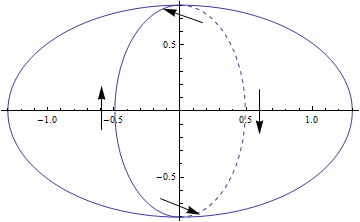}
\caption{Three external strain field directions to crack the glass. (1) An expansion or contraction; (2) pull in $e_{yy}$ strain direction while squeeze in $e_{zz}$ direction, or vice versa; (3) shear in $yz$ plane, please note $\partial u_y/\partial z$ and $\partial u_z/\partial y$ not necessarily the same for this case. }
\end{figure}
\endwidetext

Finally, to verify the existence of positive-negative transitions in glass total susceptibility, we need to sum up elastic susceptibility (Eq.(\ref{7})) and non-elastic susceptibility: $\bm{\chi}^{\rm tot}=\bm{\chi}^{\rm el}+\bm{\chi}$. Please note that the elastic susceptibility does not provide positive-negative transitions. When external strain $e_{xx}$ approaches the critical values, non-elastic susceptibility presents sharp positive-negative transitions. It becomes overwhelmingly larger than the elastic susceptibility. The positive-negative transitions of total mechanical susceptibility are determined by non-elastic susceptibility. Our theory cannot explain multiple small slips in stress-strain curve, because once a slip occurs, groups of glass molecules shift positions macroscopically. We need to rewrite the entire glass Hamiltonian after the slip happens.

\section{Discussions}
In this paper we develop a generic coupled block model in glass to prove the existence of positive-negative transitions of glass mechanical susceptibility. Our Hamiltonian contains long-wavelength phonon Hamiltonian $\hat{H}^{\rm el}$, non-elastic part of Hamiltonian $\hat{H}$, stress tensor-strain field coupling $e_{ij}\hat{T}_{ij}$, and non-elastic stress-stress interaction $\hat{V}$. We use real space renormalization procedure to set up the relation between microscopic and macroscopic non-elastic susceptibilities. The most important assumption in this paper is that the correlation function of non-elastic stress tensors are diagonal in spacial coordinates: $V^{-1}\langle \hat{T}_{ij}^{(s)}\hat{T}_{kl}^{(s')}\rangle=\chi_{ijkl}\delta_{ss'}$. With the presence of external static deformation, the spherical glass loses isotropicity, leading to the anisotropicity of non-elastic susceptibility at experimental length scale. Non-elastic susceptibility presents steep positive-negative transitions when external strain passes through critical values. However, this is still a tentative result, since it is based on the values of microscopic length scale non-elastic susceptibilities $\chi_t(L_1)$ and $\chi_l(L_1)$ which we do not know.

The basic requirement of our real space renormalization technique is that the characteristic thermal phonon wavelength $\lambda_T$ cannot exceed the starting microscopic length scale: $\lambda_T>50\AA$. Therefore, our renormalization technique is only valid below the temperature of $60$K. However,  up to now we cannot find any glass avalanche experiment below the temperature of 60K\cite{Argon2005, Gauthier2004, Dahmen2014, Fineberg1991, Fineberg1992}. We hope more experiments on the mechanical properties of amorphous materials at low temperatures could be carried out to test the validity of our theory.

\section{Acknowledgement}
D. Zhou wishes to express his deepest thanks to his advisor Anthony J. Leggett. D. Zhou also thanks Karin A. Dahmen and Xueda Wen for insightful discussions. D. Zhou would like to especially thank the anonymous referee for precious suggestions. This work is supported by the National Science Foundation under Grant No. NSF-DMR 09-06921 at the University of Illinois.

\widetext
\appendix

\section{Glass Single Block Hamiltonian $\hat{H}_0^{(s)}(\bm{e})$ with the Presence of External Static Uniform Strain $\bm{e}$}
In this subsection we try to consider glass single block Hamiltonian $\hat{H}_0^{(s)}(\bm{e})$ with the presence of external static, uniform strain field $\bm{e}$. As the simplest case, we consider the external strain as $e_{xx}=e$, $e_{yy}=e_{zz}=e_{xy}=e_{yz}=e_{zx}=0$. For other kinds of external strain ${\bm e}=e_{ij}$, similar positive-negative transition behaviors of glass mechanical susceptibility will be carried out as well.

It seems that the glass single block Hamiltonian simply adds a coupling term between external strain and non-elastic stress tensor, $\hat{H}^{(s)}_0(\bm{e})=\hat{H}^{(s)}_0+e_{xx}\hat{T}_{xx}^{(s)}$. However, one problem arises with the presence of external strain: the non-elastic stress tensor $\hat{T}_{ij}^{(s)}$ at position $\vec x_s$ might be changed by the amount of $\Delta \hat{T}_{ij}^{(s)}$. A familiar example is that external strain can change the relative position of positive and negative charges of an electric dipole moment: $\Delta p_i^{(s)}=\sum_{j}(\partial u_i^{(s)}/\partial x_j)p_j^{(s)}$. However, we have no idea how does non-elastic stress tensor change. To solve this problem, let us define a new quantity, $\hat{T}_{ij}^{(s)}(\bm{e})$, which is the summation of $\hat{T}_{ij}^{(s)}$ and $\Delta\hat{T}_{ij}^{(s)}$:
\begin{eqnarray}\label{C1}
\hat{T}_{ij}^{(s)}(\bm{e})=\hat{T}_{ij}^{(s)}+\Delta \hat{T}_{ij}^{(s)}
\end{eqnarray}
Now, if we expand the single block glass Hamiltonian $\hat{H}^{(s)}(\bm{e})$ in orders of intrinsic strain $\bm{e}^{(s)}$, the coefficient of the first order expansion is ${\delta \hat{H}^{(s)}(\bm{e})}/{\delta e_{ij}^{(s)}}=\hat{T}_{ij}^{(s)}+\Delta \hat{T}_{ij}^{(s)}$, which means up to the first order of intrinsic strain $\bm{e}^{(s)}$, the newly defined quantity $\hat{T}_{ij}^{(s)}(\bm{e})$ in Eq.(\ref{C1}) is the non-elastic stress tensor with the presence of external strain $\bm{e}$. Following this argument, the coupling term between external strain and non-elastic stress tensor is given by $ e_{xx}\hat{T}_{xx}^{(s)}(\bm{e})$. The single block zero-frequency non-elastic susceptibility, defined by ${\chi}_{ijkl}(\bm{e})=V^{-1}\langle{\delta^2  \hat{H}^{(s)}(\bm{e})}/{\delta e^{(s)}_{ij}\delta e^{(s)}_{kl}}\rangle$, is given by replacing $\hat{T}^{(s)}_{ij}$ with $\hat{T}^{(s)}_{ij}(\bm{e})$ in Eq.({\ref{11}}):

\begin{eqnarray}\label{C2}
\chi_{ijkl}(\bm{e}) & = &  \chi_{ijkl}^{\rm rel}(\bm{e})+ \chi_{ijkl}^{\rm res}(\bm{e})\nonumber \\
\chi_{ijkl}^{\rm rel}(\bm{e}) & = & \frac{\beta}{V(\bm{e})}\bigg(\sum_{nm}P_n(\bm{e})P_m(\bm{e})\langle n(\bm{e})|\hat{T}_{ij}(\bm{e})|n(\bm{e})\rangle\langle m(\bm{e})|\hat{T}_{kl}(\bm{e})|m(\bm{e})\rangle\nonumber \\
 & {} & \qquad\,\,-\sum_n P_n(\bm{e})\langle n(\bm{e})|\hat{T}_{ij}(\bm{e})|n(\bm{e})\rangle \langle n(\bm{e})|\hat{T}_{kl}(\bm{e})|n(\bm{e})\rangle\bigg)\nonumber \\
\chi_{ijkl}^{\rm res}(\bm{e}) & = & \frac{1}{V(\bm{e})}\sum_{n}\sum_{m\neq n}P_n(\bm{e})\frac{\langle n(\bm{e})|\hat{T}_{ij}(\bm{e})|m(\bm{e})\rangle\langle m(\bm{e})|\hat{T}_{kl}(\bm{e})|n(\bm{e})\rangle}{E_n(\bm{e})-E_m(\bm{e})+i\eta}\nonumber \\
 & - & \frac{1}{V(\bm{e})}\sum_{m}\sum_{n\neq m}P_m(\bm{e})\frac{\langle n(\bm{e})|\hat{T}_{ij}(\bm{e})|m(\bm{e})\rangle\langle m(\bm{e})|\hat{T}_{kl}(\bm{e})|n(\bm{e})\rangle}{E_n(\bm{e})-E_m(\bm{e})+i\eta}
\end{eqnarray}
where we define $|n(\bm{e})\rangle$ and $E_n(\bm{e})$ to be the $n$-th eigenstate and eigenvalue of single block non-elastic Hamiltonian $\hat{H}_0^{(s)}(\bm{e})$. The single block volume $V(\bm{e})=(1+e_{xx})L^3$. We define $P_n(\bm{e})=e^{-\beta E_n(\bm{e})}/\mathcal{Z}(\bm{e})$ to be the probability function of $n$-th eigenstate of $\hat{H}_0^{(s)}(\bm{e})$.

\section{Glass Super Block Hamiltonian $\hat{H}_0^{\rm super}(\bm{e})$ with the Presence of External Static Uniform Strain $\bm{e}$}
Next let us discuss glass super block Hamiltonian $\hat{H}^{\rm super}(\bm{e})$ with the presence of external strain $\bm{e}=e_{xx}$. According to the discussions in Appendix (A), glass single block Hamiltonian simply adds a coupling term, $\sum_s\hat{H}_0^{(s)}(\bm{e})=\sum_s\hat{H}_0^{(s)}+\sum_{s=1}^{N_0^3}e_{xx}\hat{T}_{xx}^{(s)}(\bm{e})$. On the other hand, it seems that non-elastic stress-stress interaction $\hat{V}(\bm{e})$ is given by $\hat{V}(\bm{e})=\sum_{ss'}\sum_{ijkl}\Lambda_{ijkl}^{(ss')}\hat{T}_{ij}^{(s)}(\bm{e})\hat{T}_{kl}^{(s')}(\bm{e})$ due to the virtual phonon exchange process.

However, the non-elastic stress-stress interaction mentioned above is not correct, because the relative positions $\vec x_s-\vec x_s'$ between different single blocks can be changed by external strain, resulting in the modification of the coefficient of non-elastic stress-stress interaction, $\Lambda_{ijkl}^{(ss')}\to \Lambda_{ijkl}^{(ss')}(\bm{e})$. Glass super block Hamiltonian is therefore given by
\begin{eqnarray}\label{D1}
\hat{H}_0^{\rm super}(\bm{e}) & = & \sum_{s}\bigg(\hat{H}_0^{(s)}+\sum_{ij}e_{xx}\hat{T}_{xx}^{(s)}(\bm{e})\bigg)+\sum_{s\neq s'}\sum_{ijkl}\Lambda_{ijkl}^{(ss')}(\bm{e})\hat{T}_{ij}^{(s)}(\bm{e})\hat{T}_{kl}^{(s')}(\bm{e})
\end{eqnarray}
We denote $|n^*(\bm{e})\rangle$ and $E_n^*(\bm{e})$ to be the $n^*$-th eigenstate and eigenvalue of $\hat{H}_0^{\rm super}(\bm{e})$, which will be useful in later discussions.

According to the definition of non-elastic stress tensor, Eq.(\ref{9}), the super block non-elastic stress tensor is defined as the derivative of $\hat{H}^{\rm super}(\bm{e})$ with respect to infinitesimal intrinsic strain: $\hat{T}_{ij}^{\rm super}(\bm{e})={\delta  \hat{H}^{\rm super}(\bm{e})}/{\delta e_{ij}}$. Because of the strain field dependence of $\Lambda_{ijkl}^{(ss')}(\bm{e})$, an extra term proportional to ${\delta \Lambda_{abcd}^{(ss')}(\bm{e})}/{\delta e_{ij}}$ appears in super block stress tensor:
\begin{eqnarray}\label{D2}
\hat{T}_{ij}^{\rm super}(\bm{e}) & = & \sum_s\hat{T}_{ij}^{(s)} (\bm{e})+\sum_{ss'}\sum_{abcd}\frac{\delta \Lambda_{abcd}^{(ss')}(\bm{e})}{\delta e_{ij}}\hat{T}_{ab}^{(s)}(\bm{e})\hat{T}_{cd}^{(s')}(\bm{e})
\end{eqnarray}

Smilar with super block non-elastic stress tensor, the super block non-elastic susceptibility also receives an extra term. According to the definition of non-elastic susceptibility Eq.(\ref{4}), to calculate super block non-elastic susceptibility with the presence of external strain $\bm{e}$, we have to induce an infinitesimal extra external strain $e_{ij}(t)=e_{ij}\left(e^{-i\omega t}+e^{+i\omega t}\right)$, with wavelength much greater than the side of super block of glass: $\frac{2\pi}{k}\gg N_0L$, so that the Hamiltonian $\hat{H}^{\rm super}(\bm{e})$ provides a corresponding extra stress response $\langle \hat{T}_{ij}^{\rm super}(\bm{e})\rangle(t)=(\langle \hat{T}_{ij}^{\rm super}(\bm{e})\rangle e^{-i\omega t}+c.c)$. The ``stress response of super block non-elastic Hamiltonian", $\langle \hat{T}_{ij}^{\rm super}(\bm{e})\rangle(t)$, is by definition given as follows

\begin{eqnarray}\label{D3}
\langle \hat{T}_{ij}^{\rm super}(\bm{e})\rangle (t)=\sum_{m^*}\frac{e^{-\beta E_m^*(\bm{e})}}{\mathcal{Z}^*(\bm{e})}\langle m^*(\bm{e}), t|\hat{T}_{ij}^{\rm super}(\bm{e})|m^*(\bm{e}), t\rangle\quad\quad
\end{eqnarray}
where $|m^*(\bm{e}),t\rangle$ is the $m^*$-th eigenstate wave function of super block non-elastic Hamiltonian $\hat{H}^{\rm super}(\bm{e})$ perturbed by infinitesimal perturbation $\sum_{ij}e_{ij}(t)\hat{T}_{ij}^{\rm super}(\bm{e})$, and $\mathcal{Z}^*(\bm{e})=\sum_{m^*}e^{-\beta E_m^*(\bm{e})}$ is the partition function of $\hat{H}^{\rm super}(\bm{e})$. By using linear response theory we expand $\langle \hat{T}_{ij}^{\rm super}(\bm{e})\rangle(t)$ up to the first order of perturbation $\sum_{ij}e_{ij}(t)\hat{T}_{ij}^{\rm super}(\bm{e})$, to get the super block non-elastic susceptibility below. 
\begin{eqnarray}\label{D4}
\chi_{ijkl}^{\rm super}(\bm{e})
 & = & \frac{\beta}{N_0^3V(\bm{e})}\bigg(\sum_{n^*m^*}P_n^*(\bm{e})P_m^*(\bm{e})\langle n^*(\bm{e})|\sum_s\hat{T}^{(s)}_{ij}(\bm{e})|n^*(\bm{e})\rangle\langle m^*(\bm{e})|\sum_{s'}\hat{T}^{(s')}_{kl}(\bm{e})|m^*(\bm{e})\rangle\nonumber \\
 & {} & \qquad\qquad -\sum_{n} P_n^*(\bm{e})\langle n^*(\bm{e})|\sum_{s}\hat{T}^{(s)}_{ij}(\bm{e})|n^*(\bm{e})\rangle \langle n^*(\bm{e})|\sum_{s'}\hat{T}^{(s')}_{kl}(\bm{e})|n^*(\bm{e})\rangle\bigg)\nonumber \\
 & + & \frac{1}{N_0^3V(\bm{e})}\bigg(\sum_{n^*}\sum_{m^*\neq n^*}P_n^*(\bm{e})\frac{\langle n^*(\bm{e})|\sum_s\hat{T}^{(s)}_{ij}(\bm{e})|m^*(\bm{e})\rangle\langle m^*(\bm{e})|\sum_{s'}\hat{T}^{(s')}_{kl}(\bm{e})|n^*(\bm{e})\rangle}{E_n^*(\bm{e})-E^*_m(\bm{e})+i\eta}\nonumber \\
 & {} & \qquad\qquad -\sum_{m^*}\sum_{n^*\neq m^*}P_m^*(\bm{e})\frac{\langle n^*(\bm{e})|\sum_s\hat{T}^{(s)}_{ij}(\bm{e})|m^*(\bm{e})\rangle\langle m^*(\bm{e})|\sum_{s'}\hat{T}^{(s')}_{kl}(\bm{e})|n^*(\bm{e})\rangle}{E_n^*(\bm{e})-E^*_m(\bm{e})+i\eta}\bigg)\nonumber \\
 & + & \frac{1}{N_0^3V(\bm{e})}\sum_{abcd}\sum_{ss'}\left<\frac{\delta^2\Lambda_{abcd}^{(ss')}(\bm{e})}{\delta e_{ij}\delta e_{kl}}\hat{T}_{ab}^{(s)}(\bm{e})\hat{T}_{cd}^{(s')}(\bm{e})\right>
\end{eqnarray}
In the above result, $P_n^*(\bm{e})=e^{-\beta E_n^*(\bm{e})}/\mathcal{Z}^*(\bm{e})$ is the probability function of the $n^*$-th eigenstate. The first and second lines stand for super block zero-frequency relaxation and resonance susceptibilities, while the third line is the extra term in super block non-elastic susceptibility which comes from the second term of Eq.(\ref{D2}).

\section{Derivations of Renormalization Equation of Non-elastic Stress-Stress Susceptibility}
In this appendix we want to give a detailed derivation in obtaining the real space renormalization equation of non-elastic stress-stress susceptibility, Eq.(\ref{23}). That is, we want to set up the relation between super block non-elastic susceptibility Eq.(\ref{D4}) and single block non-elastic susceptibility Eq.(\ref{C2}). For notation simplicity, in this section we use $|n\rangle$, $E_n$ to represent $|n(\bm{e})\rangle$, $E_n(\bm{e})$, and use $|n^*\rangle$, $E_n^*$ to represent $|n^*(\bm{e})\rangle$, $E_n^*(\bm{e})$. We also use $\hat{V}$, $V$, $\hat{T}_{ij}$, $\hat{T}_{ij}^{\rm super}$, $\chi_{ijkl}$ and $\chi_{ijkl}^{\rm super}$ to represent $\hat{V}(\bm{e})$(non-elastic stress-stress interaction), $V(\bm{e})$(the volume of single block glass), $\hat{T}_{ij}(\bm{e})$, $\hat{T}_{ij}^{\rm super}(\bm{e})$, $\chi_{ijkl}(\bm{e})$ and $\chi_{ijkl}^{\rm super}(\bm{e})$. Such simplification does not affect the final result, Eq.(\ref{23}). We treat non-elastic stress-stress interaction $\hat{V}(\bm{e})$ as perturbation. By using time-independent perturbation theory, we obtain the following relations between $|n\rangle$, $E_n$ and $|n^*\rangle$, $E_n^*$
\begin{eqnarray}\label{B2}
|n^*\rangle = |n\rangle+\sum_{p\neq n}\frac{\langle p|\hat{V}|n\rangle}{E_n-E_p}|p\rangle+\mathcal{O}(V^2)\quad\quad\quad\quad 
E_n^* = E_n+\langle n|\hat{V}|n\rangle +\mathcal{O}(V^2)
\end{eqnarray}
We expand the super block partition function and probability function up to the first order in $\hat{V}$: 
\begin{eqnarray}\label{B3}
e^{-\beta E_n^*} = e^{-\beta E_n}\left(1-\beta\langle n|\hat{V}|n\rangle +\mathcal{O}(V^2)\right)\quad\quad\quad
\mathcal{Z}^* = \sum_le^{-\beta E_l}\left(1-\beta\langle l|\hat{V}|l\rangle +\mathcal{O}(V^2)\right)
\end{eqnarray}
Let us denote 
\begin{eqnarray}\label{E1}
\delta|n\rangle = \sum_{p\neq n}\frac{\langle p|\hat{V}|n\rangle}{E_n-E_p}|p\rangle\qquad\quad 
\delta E_n = \langle n|\hat{V}|n\rangle\qquad\quad
\delta\mathcal{Z} = -\sum_le^{-\beta E_l}\beta\langle l|\hat{V}|l\rangle 
\end{eqnarray}
to represent the first order expansions of the eigenstates, eigenvalues and partition functions. The following definitions will be very useful in details of calculations:
\begin{eqnarray}\label{B4}
\chi_{ijkl}^{\rm rel(1)}
 & = & 
\frac{1}{V}\beta\sum_{nm}P_nP_m\langle n|\hat{T}_{ij}|n\rangle \langle m|\hat{T}_{kl}|m\rangle\nonumber \\
\chi_{ijkl}^{\rm rel(2)}
 & = & 
\frac{1}{V}\beta\sum_nP_n\langle n|\hat{T}_{ij}|n\rangle \langle n|\hat{T}_{kl}|n\rangle  
\nonumber \\
\chi_{ijkl}^{\rm res}
 & = & 
\frac{2}{V\hbar}\sum_{nl}\frac{e^{-\beta E_n}}{\mathcal{Z}}\langle l|\hat{T}_{ij}|n\rangle\langle n|\hat{T}_{kl}|l\rangle\frac{\omega_{ln}}{(i\eta)^2-\omega_{ln}^{2}}
\end{eqnarray}
where $\omega_{ln}=(E_l-E_n)/\hbar$. Therefore the non-elastic susceptibility is written as follows,
\begin{eqnarray}\label{B5}
\chi_{ijkl}
 & = & 
\left(\chi_{ijkl}^{\rm rel(1)}-\chi_{ijkl}^{\rm rel(2)} \right)
+\chi_{ijkl}^{\rm res}
\end{eqnarray}
In the rest of this appendix we want to expand three parts of super block non-elastic susceptibility, $\chi_{ijkl}^{\rm super \, rel(1)}$, $\chi_{ijkl}^{\rm super \, rel(2)}$ and $\chi_{ijkl}^{\rm super \,res}$ up to the first order of interaction $\hat{V}$ (i.e., the second order of single block susceptibility).

\subsection{Expansion details for $\chi_{ijkl}^{\rm super \,rel(1)}$}
\begin{eqnarray}\label{B6}
\chi_{ijkl}^{\rm super \,rel(1)} & = & 
\frac{\beta}{N_0^3V}\sum_{n^*m^*}\frac{e^{-\beta (E_n^*+E_m^*)}}{\mathcal{Z}^{*2}}\langle n^*|\hat{T}_{ij}|n^*\rangle \langle m^*|\hat{T}_{kl}|m^*\rangle  \nonumber \\
 & = & 
\frac{\beta}{N_0^3V}\sum_{nm}\frac{e^{-\beta (E_n+E_m)}}{\mathcal{Z}^2}\langle n|\hat{T}_{ij}|n\rangle \langle m|\hat{T}_{kl}|m\rangle  \nonumber \\
 & + & \frac{\beta}{N_0^3V}\sum_{nm}\frac{e^{-\beta(E_n+E_m)}(-\beta \delta E_n-\beta\delta E_m)}{\mathcal{Z}^2}\langle n|\hat{T}_{ij}|n\rangle\langle m|\hat{T}_{kl}|m\rangle
\qquad\qquad\qquad\qquad J_1\nonumber \\
 & + & \frac{\beta}{N_0^3V}\sum_{nm}\frac{e^{-\beta(E_n+E_m)}(-2\delta\mathcal{Z})}{\mathcal{Z}^3}\langle n|\hat{T}_{ij}|n\rangle\langle m|\hat{T}_{kl}|m\rangle\qquad\qquad\qquad\qquad\qquad\qquad\,\,\, J_2\nonumber \\
 & + & \frac{\beta}{N_0^3V}\sum_{nm}\frac{e^{-\beta(E_n+E_m)}}{\mathcal{Z}^2} \bigg[\left(\delta\langle n|\right)\hat{T}_{ij}|n\rangle\langle m|\hat{T}_{kl}|m\rangle+\langle n|\hat{T}_{ij}\left(\delta|n\rangle\right)\langle m|\hat{T}_{kl}|m\rangle\nonumber \\
 & {} & \qquad\qquad\qquad\qquad\qquad+\langle n|\hat{T}_{ij}|n\rangle\left(\delta\langle m|\right)\hat{T}_{kl}|m\rangle+\langle n|\hat{T}_{ij}|n\rangle\langle m|\hat{T}_{kl}\left(\delta|m\rangle\right)\bigg]\quad\,\,\,\, J_3
\end{eqnarray}
where $\delta|n\rangle$, $\delta\mathcal{Z}$ and $\delta E_n$ stand for the first order expansions defined in Eq.(\ref{E1}). Now we begin to calculate every expansion terms $J_1, J_2, J_3$ in the above result. \\
Expansion for term $J_1$:
\begin{eqnarray}\label{B6.1}
J_1 & = & -\frac{\beta^2}{N_0^3V}\sum_{nm}\frac{e^{-\beta(E_n+E_m)}( \delta E_n+\delta E_m)}{\mathcal{Z}^2}\langle n|\hat{T}_{ij}|n\rangle\langle m|\hat{T}_{kl}|m\rangle\nonumber \\
 & = & -\frac{\beta^2}{N_0^3V}\sum_{nm}\frac{e^{-\beta(E_n+E_m)}}{\mathcal{Z}^2}\left(\langle n|\hat{V}|n\rangle+\langle m|\hat{V}|m\rangle\right)\langle n|\hat{T}_{ij}|n\rangle\langle m|\hat{T}_{kl}|m\rangle\nonumber \\
 & = & -\frac{\beta^2}{N_0^3V}\sum_{nm}\frac{e^{-\beta(E_n+E_m)}}{\mathcal{Z}^2}
\sum_{abcd}\sum_{uu'}\sum_{ss'}\Lambda_{abcd}^{(uu')}
\left(\langle n|\hat{T}_{ab}^{(u)}\hat{T}_{cd}^{(u')}|n\rangle+\langle m|\hat{T}_{ab}^{(u)}\hat{T}_{cd}^{(u')}|m\rangle\right)\langle n|\hat{T}^{(s)}_{ij}|n\rangle\langle m|\hat{T}^{(s')}_{kl}|m\rangle\nonumber \\
 & = & -\frac{\beta^2}{N_0^3V}\sum_{nm}\frac{e^{-\beta(E_n+E_m)}}{\mathcal{Z}^2}
\sum_{abcd}\sum_{uu'}\sum_{ss'}\Lambda_{abcd}^{(uu')}\nonumber \\
 & {} & 
\left(\langle n|\hat{T}_{ab}^{(u)}\sum_{l}|l\rangle\langle l|\hat{T}_{cd}^{(u')}|n\rangle+\langle m|\hat{T}_{ab}^{(u)}\sum_{l}|l\rangle\langle l|\hat{T}_{cd}^{(u')}|m\rangle\right)\langle n|\hat{T}^{(s)}_{ij}|n\rangle\langle m|\hat{T}^{(s')}_{kl}|m\rangle
\end{eqnarray}
Let's stop here for a moment and talk about how could we write the above result in terms of single block susceptibility. Please note that we have only defined the single block relaxation and resonance susceptibilities, Eq.(\ref{B4}). The relaxation susceptibility (part 1 and part 2 of relaxation susceptibilities in Eq.(\ref{B4})) is the product of diagonal matrix elements of stress tensors; the resonance susceptibility is the product of off-diagonal matrix elements of stress tensors. So the question is, why do we never define such a term, that is the product between diagonal and off-diagonal matrix element of stress tensors? 

The reason is after averaging over spacial coordinate $\vec x_S$, such kind of product between diagonal and off-diagonal matrix elements of stress tensors $\hat{T}_{IJ}^{(S)}$ will vanish, becasue the diagonal and off-diagonal $\hat{T}_{IJ}^{(S)}$ matrix elements are random as the function of spacial coordinate $\vec x_S$. In other words, there is no specific relation between diagonal and off-diagonal matrix elements. In addition, the diagonal matrix element $\langle N|\hat{T}_{IJ}^{(S)}|N\rangle\propto \delta\langle\hat{H}^{\rm non}\rangle/\delta e_{IJ}$ is defined as the ``non-elastic" stress tensor in glass. It is highly plausible that the non-elastic stress tensor expectation value tends to vanish for large enough block of glass.

There is another problem for the pairing rule of stress tensor matrix elements: can we pair matrix elements between different blocks $\vec x_S\neq \vec x_S'$? For example, does the term $\langle N|\hat{T}_{IJ}^{(S)}|M\rangle \langle M|\hat{T}_{KL}^{(S')}|N\rangle$ with $S\neq S'$ vanish or not? Again, because the diagonal and off-diagonal stress tensor matrix elements are random as the function of spacial coordinate $\vec x_S$, after the summation over $S, S'$, $\sum_{SS'}\langle N|\hat{T}_{IJ}^{(S)}|M\rangle \langle M|\hat{T}_{KL}^{(S')}|N\rangle$ turns out to be zero. Therefore, the matrix element of stress tensor at the $S$-th block must be paired with the matrix element at the same $S$-th block. In other words, there is no obvious relation between stress tensors at different blocks.

Based on the above three reasons, we obtain the following two rules of matrix element pairing: suppose we have, for example, a diagonal matrix element $\langle N|\hat{T}_{IJ}^{(S)}|N\rangle$ and an off-diagonal matrix element $\langle N|\hat{T}_{IJ}^{(S)}|M\rangle$. The diagonal matrix element at the $S$-th block, $\langle N|\hat{T}_{IJ}^{(S)}|N\rangle$, is required to be paired with the diagonal matrix element of stress tensor at the same $S$-th block; the off-diagonal matrix element at the $S$-th block, $\langle N|\hat{T}_{IJ}^{(S)}|M\rangle$ is required to be pair with the off-diagonal matrix element at the same $S$-th block.

Now let us go back to the final result of Eq.(\ref{B6.1}). There are two summations: 
\begin{eqnarray}\label{B6.2}
 & {} & \sum_{uu'}\sum_{ss'}\Lambda_{abcd}^{(uu')}
\langle n|\hat{T}_{ab}^{(u)}\sum_{l}|l\rangle\langle l|\hat{T}_{cd}^{(u')}|n\rangle\langle n|\hat{T}^{(s)}_{ij}|n\rangle\langle m|\hat{T}^{(s')}_{kl}|m\rangle\nonumber \\
 & {} & \sum_{uu'}\sum_{ss'}\Lambda_{abcd}^{(uu')}
\langle m|\hat{T}_{ab}^{(u)}\sum_{l}|l\rangle\langle l|\hat{T}_{cd}^{(u')}|m\rangle\langle n|\hat{T}^{(s)}_{ij}|n\rangle\langle m|\hat{T}^{(s')}_{kl}|m\rangle
\end{eqnarray}
The coefficient of non-elastic stress-stress interaction, $\Lambda_{abcd}^{(uu')}$, does not allow $u=u'$. Therefore, the matrix elements $\langle n|\hat{T}_{ab}^{(u)}|l\rangle$, $\langle l|\hat{T}_{cd}^{(u')}|n\rangle$, and $\langle m|\hat{T}_{ab}^{(u)}|l\rangle$, $\langle l|\hat{T}_{cd}^{(u')}|m\rangle$, are not allowed to pair with each other. In the first summation of Eq.(\ref{B6.2}), the only two possible ways of pairing is:
\begin{eqnarray}\label{B6.3}
 & {} & \langle n|\hat{T}_{ab}^{(u)}|l\rangle \quad {\rm paired \,\,\, with}\quad \langle n|\hat{T}^{(s)}_{ij}|n\rangle; \quad\quad
\langle l|\hat{T}_{cd}^{(u')}|n\rangle \quad {\rm paired \,\,\, with}\quad \langle m|\hat{T}^{(s')}_{kl}|m\rangle\nonumber \\
 & {\rm or\quad} & 
\langle n|\hat{T}_{ab}^{(u)}|l\rangle \quad {\rm paired \,\,\, with}\quad \langle m|\hat{T}^{(s')}_{kl}|m\rangle; \quad\quad
\langle l|\hat{T}_{cd}^{(u')}|n\rangle \quad {\rm paired \,\,\, with}\quad \langle n|\hat{T}^{(s)}_{ij}|n\rangle
\end{eqnarray}
the first pairing candidate requires $u=s, u'=s'$, and $|l\rangle=|n\rangle$; the second pairing candidate requires $u=s', u'=s$, and $|l\rangle=|n\rangle$. In the second summation of Eq.(\ref{B6.2}) the possible pairings require $u=s, u'=s'$ or $u=s', u'=s$, and $|l\rangle=|m\rangle$. With the above pairing rules, finally we can procede the calculation of term $J_1$ from Eq.(\ref{B6.1}) as follows,

\begin{eqnarray}\label{B7}
J_1 & = & -\frac{\beta^2}{N_0^3V}
\sum_{abcd}\sum_{ss'}\sum_{n^{(s)}n^{(s')}m^{(s')}}\frac{e^{-\beta(E_n^{(s)}+E_n^{(s')}+E_m^{(s')})}}{\mathcal{Z}^{(s)}\mathcal{Z}^{(s')2}}\Lambda_{abcd}^{(ss')}e^{-ik(x_s-x_s')}\nonumber \\
 & {} & \bigg\{\langle n^{(s)}|\hat{T}_{cd}^{(s)}|n^{(s)}\rangle\langle n^{(s)}|\hat{T}_{ij}^{(s)}|n^{(s)}\rangle\langle m^{(s')}|\hat{T}_{kl}^{(s')}|m^{(s')}\rangle\langle n^{(s')}|\hat{T}_{ab}^{(s')}|n^{(s')}\rangle \nonumber \\
 & {} & +
\langle n^{(s)}|\hat{T}_{ij}^{(s)}|n^{(s)}\rangle\langle m^{(s)}|\hat{T}_{ab}^{(s)}|m^{(s)}\rangle\langle m^{(s')}|\hat{T}_{cd}^{(s')}|m^{(s')}\rangle\langle m^{(s')}|\hat{T}_{kl}^{(s')}|m^{(s')}\rangle 
\bigg\}\nonumber \\
 & = & -\frac{V}{N_0^3}\sum_{abcd}\sum_{ss}\Lambda_{abcd}^{(ss')}e^{-ik(x_s-x_s')}\left(\chi_{cdij}^{\rm rel(2)}\chi_{abkl}^{\rm rel(1)}+\chi_{cdkl}^{\rm rel(2)}\chi_{abij}^{\rm rel(1)}\right)
\end{eqnarray}
Next, we consider the expansion for term $J_2$:
\begin{eqnarray}\label{B8}
J_2 & = & -\frac{2\beta}{N_0^3V}\sum_{nm}\frac{e^{-\beta(E_n+E_m)}}{\mathcal{Z}^3}\delta\mathcal{Z}\langle n|\hat{T}_{ij}|n\rangle \langle m|\hat{T}_{kl}|m\rangle\nonumber \\
 & = & \frac{2\beta^2}{N_0^3V}\sum_{lmn}\frac{e^{-\beta(E_n+E_m+E_l)}}{\mathcal{Z}^3}\sum_{abcd}\sum_{uu'}\sum_{ss'}\Lambda_{abcd}^{(uu')}\langle l|\hat{T}_{ab}^{(u)}\hat{T}_{cd}^{(u')}|l\rangle\langle n|\hat{T}_{ij}^{(s)}|n\rangle\langle m|\hat{T}_{kl}^{(s')}|m\rangle\nonumber \\
 & = & \frac{2\beta^2}{N_0^3V}\sum_{lmn}\frac{e^{-\beta(E_n+E_m+E_l)}}{\mathcal{Z}^3}\sum_{abcd}\sum_{uu'}\sum_{ss'}\Lambda_{abcd}^{(uu')}\langle l|\hat{T}_{ab}^{(u)}\sum_k|k\rangle\langle k|\hat{T}_{cd}^{(u')}|l\rangle\langle n|\hat{T}_{ij}^{(s)}|n\rangle\langle m|\hat{T}_{kl}^{(s')}|m\rangle\nonumber \\
 & = & \frac{2\beta^2}{N_0^3V}\sum_{l^{(s)}l^{(s')}m^{(s')}n^{(s)}}\frac{e^{-\beta(E_n^{(s)}+E_m^{(s')}+E_l^{(s)}+E_l^{(s')})}}{\mathcal{Z}^{(s)2}\mathcal{Z}^{(s')2}}\sum_{abcd}\sum_{ss'}\Lambda_{abcd}^{(ss')}\nonumber \\
 & {} & \langle l^{(s')}|\hat{T}_{cd}^{(s')}|l^{(s')}\rangle\langle m^{(s')}|\hat{T}_{kl}^{(s')}|m^{(s')}\rangle\langle n^{(s)}|\hat{T}_{ij}^{(s)}|n^{(s)}\rangle\langle l^{(s)}|\hat{T}_{ab}^{(s)}|l^{(s)}\rangle\nonumber \\
 & = & \frac{2V}{N_0^3}\sum_{abcd}\sum_{ss}\Lambda_{abcd}^{(ss')}\chi_{cdij}^{\rm rel(1)}\chi_{abkl}^{\rm rel(1)}
\end{eqnarray}
In the above calculations, because the coefficient $\Lambda_{abcd}^{(uu')}$ does not allow $u=u'$, the matrix elements $\langle l|\hat{T}_{ab}^{(u)}|k\rangle$, $\langle k|\hat{T}_{cd}^{(u')}|l\rangle$ are not allowed to pair with each other. We need to pair matrix elements $\langle l|\hat{T}_{ab}^{(u)}|k\rangle$, $\langle k|\hat{T}_{cd}^{(u')}|l\rangle$ with diagonal matrix elements $\langle n|\hat{T}_{ij}^{(s)}|n\rangle$, $\langle m|\hat{T}_{kl}^{(s')}|m\rangle$. Since diagonal matrix elements are only allowed to be paired with diagonal matrix elements, the choice of quantum number $k$ has to be $k=l$, so that $\langle l|\hat{T}_{ab}^{(u)}|k\rangle$ is a diagonal matrix element. The diagonal matrix element $\langle l|\hat{T}_{ab}^{(u)}|k=l\rangle$ can be paired with $\langle n|\hat{T}_{ij}^{(s)}|n\rangle$ or $\langle m|\hat{T}_{kl}^{(s')}|m\rangle$. Finally we obtain the result in Eq.(\ref{B8}).
\\
The expansion for term $J_3$:
\begin{eqnarray}\label{B9.1}
J_3 & = & \frac{\beta}{N_0^3V}\sum_{nm}\frac{e^{-\beta (E_n+E_m)}}{\mathcal{Z}^2}\bigg[\left(\delta\langle n|\right)\hat{T}_{ij}|n\rangle\langle m|\hat{T}_{kl}|m\rangle+\langle n|\hat{T}_{ij}\left(\delta|n\rangle \right)\langle m|\hat{T}_{kl}|m\rangle\nonumber \\
 & {} & \qquad\qquad\qquad\qquad\qquad+\langle n|\hat{T}_{ij}|n\rangle\left(\delta\langle m|\right)\hat{T}_{kl}|m\rangle+\langle n|\hat{T}_{ij}|n\rangle\langle m|\hat{T}_{kl}\left(\delta|m\rangle\right)\bigg]\nonumber \\
 & = & \frac{\beta}{N_0^3V}\sum_{lmn}\sum_{abcd}\sum_{uu'}\sum_{ss'}\frac{1}{E_n-E_l}\frac{e^{-\beta (E_n+E_m)}}{\mathcal{Z}^2}\Lambda_{abcd}^{(uu')}\nonumber \\
 & {} & \left(\langle n|\hat{T}_{ab}^{(u)}\hat{T}_{cd}^{(u')}|l\rangle\langle l|\hat{T}_{ij}^{(s)}|n\rangle\langle m|\hat{T}_{kl}^{(s')}|m\rangle+\langle n|\hat{T}_{ij}^{(s)}|l\rangle \langle l|\hat{T}_{ab}^{(u)}\hat{T}_{cd}^{(u')}|n\rangle \langle m|\hat{T}_{kl}^{(s')}|m\rangle\right)\nonumber \\
 & + & \frac{\beta}{N_0^3V}\sum_{lmn}\sum_{abcd}\sum_{uu'}\sum_{ss'}\frac{1}{E_m-E_l}\frac{e^{-\beta (E_n+E_m)}}{\mathcal{Z}^2}\Lambda_{abcd}^{(uu')}\nonumber \\
 & {} & \left(\langle n|\hat{T}_{ij}^{(s)}|n\rangle \langle m|\hat{T}_{ab}^{(u)}\hat{T}_{cd}^{(u')}|l\rangle \langle l|\hat{T}_{kl}^{(s')}|m\rangle +\langle n|\hat{T}_{ij}^{(s)}|n\rangle \langle m|\hat{T}_{kl}^{(s')}|l\rangle\langle l|\hat{T}_{ab}^{(u)}\hat{T}_{cd}^{(u')}|m\rangle\right)
\end{eqnarray}
At this stage we insert the identity between $\hat{T}_{ab}^{(u)}$ and $\hat{T}_{cd}^{(u')}$: $\sum_k|k\rangle \langle k|$. 
\begin{eqnarray}\label{B9}
J_3 & = & \frac{\beta}{N_0^3V}\sum_{lmn}\sum_{abcd}\sum_{uu'}\sum_{ss'}\frac{1}{E_n-E_l}\frac{e^{-\beta (E_n+E_m)}}{\mathcal{Z}^2}\Lambda_{abcd}^{(uu')}\nonumber \\
 & {} & \left(\langle n|\hat{T}_{ab}^{(u)}\sum_k|k\rangle\langle k|\hat{T}_{cd}^{(u')}|l\rangle\langle l|\hat{T}_{ij}^{(s)}|n\rangle\langle m|\hat{T}_{kl}^{(s')}|m\rangle+\langle n|\hat{T}_{ij}^{(s)}|l\rangle \langle l|\hat{T}_{ab}^{(u)}\sum_k|k\rangle\langle k|\hat{T}_{cd}^{(u')}|n\rangle \langle m|\hat{T}_{kl}^{(s')}|m\rangle\right)\nonumber \\
 & + & \frac{\beta}{N_0^3V}\sum_{lmn}\sum_{abcd}\sum_{uu'}\sum_{ss'}\frac{1}{E_m-E_l}\frac{e^{-\beta (E_n+E_m)}}{\mathcal{Z}^2}\Lambda_{abcd}^{(uu')}\nonumber \\
 & {} & \left(\langle n|\hat{T}_{ij}^{(s)}|n\rangle \langle m|\hat{T}_{ab}^{(u)}\sum_k|k\rangle\langle k|\hat{T}_{cd}^{(u')}|l\rangle \langle l|\hat{T}_{kl}^{(s')}|m\rangle +\langle n|\hat{T}_{ij}^{(s)}|n\rangle \langle m|\hat{T}_{kl}^{(s')}|l\rangle\langle l|\hat{T}_{ab}^{(u)}\sum_k|k\rangle\langle k|\hat{T}_{cd}^{(u')}|m\rangle\right)\nonumber \\
 & = & \frac{2\beta}{N_0^3V}\sum_{abcd}\sum_{ss'}\sum_{l^{(s)}m^{(s')}n^{(s)}n^{(s')}}\frac{e^{-\beta (E_n^{(s)}+E_n^{(s')}+E_m^{(s')})}}{\mathcal{Z}^{(s)}\mathcal{Z}^{(s')2}}\Lambda_{abcd}^{(ss')}\nonumber \\
 & {} & \langle m^{(s')}|\hat{T}_{kl}^{(s')}|m^{(s')}\rangle \langle n^{(s')}|\hat{T}_{cd}^{(s')}|n^{(s')}\rangle\frac{\langle l^{(s)}|\hat{T}_{ij}^{(s)}|n^{(s)}\rangle\langle n^{(s)}|\hat{T}_{ab}^{(s)}|l^{(s)}\rangle}{E_n^{(s)}-E_l^{(s)}+i\eta}\nonumber \\
 & + & \frac{2\beta}{N_0^3V}\sum_{abcd}\sum_{ss'}\sum_{l^{(s')}m^{(s')}n^{(s)}m^{(s)}}\frac{e^{-\beta (E_n^{(s)}+E_m^{(s)}+E_m^{(s')})}}{\mathcal{Z}^{(s)2}\mathcal{Z}^{(s')}}\Lambda_{abcd}^{(ss')}
\nonumber \\
 & {} &
\langle n^{(s)}|\hat{T}_{ij}^{(s)}|n^{(s)}\rangle \langle m^{(s)}|\hat{T}_{ab}^{(s)}|m^{(s)}\rangle\frac{\langle m^{(s')}|\hat{T}_{kl}^{(s')}|l^{(s')}\rangle\langle l^{(s')}|\hat{T}_{cd}^{(s')}|m^{(s')}\rangle}{E_m^{(s')}-E_l^{(s')}+i\eta}\nonumber \\
 & = & \frac{V}{N_0^3}\sum_{abcd}\sum_{ss'}\Lambda_{abcd}^{(ss')}\left(\chi_{cdkl}^{\rm rel(1)}\chi_{abij}^{\rm res}+\chi_{abij}^{\rm rel(1)}\chi_{cdkl}^{\rm res}\right)
\end{eqnarray}
In the above calculations, coefficient $\Lambda_{abcd}^{(uu')}$ does not allow $u= u'$. In the first step of Eq.(\ref{B9}) calculations, we have 4 summations:
\begin{eqnarray}\label{B9.2}
 & {} & \sum_{lmnk}\sum_{uu'}\sum_{ss'}\Lambda_{abcd}^{(uu')}\left(\langle n|\hat{T}_{ab}^{(u)}|k\rangle\langle k|\hat{T}_{cd}^{(u')}|l\rangle\langle l|\hat{T}_{ij}^{(s)}|n\rangle\langle m|\hat{T}_{kl}^{(s')}|m\rangle+\langle n|\hat{T}_{ij}^{(s)}|l\rangle \langle l|\hat{T}_{ab}^{(u)}|k\rangle\langle k|\hat{T}_{cd}^{(u')}|n\rangle \langle m|\hat{T}_{kl}^{(s')}|m\rangle\right)\nonumber \\
 & {} & \sum_{lmnk}\sum_{uu'}\sum_{ss'}\Lambda_{abcd}^{(uu')}
\left(\langle n|\hat{T}_{ij}^{(s)}|n\rangle \langle m|\hat{T}_{ab}^{(u)}|k\rangle\langle k|\hat{T}_{cd}^{(u')}|l\rangle \langle l|\hat{T}_{kl}^{(s')}|m\rangle +\langle n|\hat{T}_{ij}^{(s)}|n\rangle \langle m|\hat{T}_{kl}^{(s')}|l\rangle\langle l|\hat{T}_{ab}^{(u)}|k\rangle\langle k|\hat{T}_{cd}^{(u')}|m\rangle\right)
\end{eqnarray}
For example, we discuss the pairing rule of the first summation only. The pairing rule for the other three summations is the same. In the first summation, the matrix elements $\langle n|\hat{T}_{ab}^{(u)}|k\rangle$, $\langle k|\hat{T}_{cd}^{(u')}|l\rangle$ cannot be paired with each other. Therefore, we need to pair $\langle n|\hat{T}_{ab}^{(u)}|k\rangle$, $\langle k|\hat{T}_{cd}^{(u')}|l\rangle$ with matrix elements $\langle l|\hat{T}_{ij}^{(s)}|n\rangle$, $\langle m|\hat{T}_{kl}^{(s')}|m\rangle$. There are two candidates: first, $\langle n|\hat{T}_{ab}^{(u)}|k\rangle$ is paired with $\langle l|\hat{T}_{ij}^{(s)}|n\rangle$, and $\langle k|\hat{T}_{cd}^{(u')}|l\rangle$ is paired with $\langle m|\hat{T}_{kl}^{(s')}|m\rangle$; second, $\langle n|\hat{T}_{ab}^{(u)}|k\rangle$ is paired with $\langle m|\hat{T}_{kl}^{(s')}|m\rangle$, and $\langle k|\hat{T}_{cd}^{(u')}|l\rangle$ is paired with $\langle l|\hat{T}_{ij}^{(s)}|n\rangle$. The first candidate forces $u=s, u'=s'$. Since the matrix element $\langle l|\hat{T}_{ij}^{(s)}|n\rangle$ is off-diagonal, $\langle n|\hat{T}_{ab}^{(u)}|k\rangle$ has to be off-diagonal as well. Therefore we choose $k=l$. In the second candidate forces $u=s', u'=s$. We also need to choose $k=n$ so that $\langle n|\hat{T}_{ab}^{(u)}|k\rangle$ is diagonal. Repeat the same process for the other three summations in Eq.(\ref{B9.2}), we finally obtain the result in Eq.(\ref{B9}).

\subsection{Expansion details for $\chi_{ijkl}^{\rm super \,rel(2)}$}
\begin{eqnarray}\label{B10}
\chi_{ijkl}^{\rm super \,rel(2)} & = & -\frac{\beta }{N_0^3V}\sum_{n^*}\frac{e^{-\beta E_n^*}}{\mathcal{Z}^*}\langle n^*|\hat{T}_{ij}|n^*\rangle\langle n^*|\hat{T}_{kl}|n^*\rangle\nonumber \\
 & = & -\frac{\beta }{N_0^3V}\sum_{n}\frac{e^{-\beta E_n}(1-\beta E_n)}{\mathcal{Z}+\delta \mathcal{Z}}\left(\langle n|+\delta \langle n|\right)\hat{T}_{ij}\left(|n\rangle+\delta|n\rangle\right)\left(\langle n|+\delta\langle n|\right)\hat{T}_{kl}\left(|n\rangle+\delta|n\rangle\right)\nonumber \\
 & = & -\frac{\beta }{N_0^3V}\sum_{n}\frac{e^{-\beta E_n}}{\mathcal{Z}}\langle n|\hat{T}_{ij}|n\rangle\langle n|\hat{T}_{kl}|n\rangle\nonumber \\
 & {} & -\frac{\beta}{N_0^3V}\sum_n\frac{e^{-\beta E_n}(-\beta \delta E_n)}{\mathcal{Z}}\langle n|\hat{T}_{ij}|n\rangle \langle n|\hat{T}_{kl}|n\rangle \nonumber \qquad\qquad\qquad\qquad\qquad\qquad J_4\\
 & {} & -\frac{\beta }{N_0^3V}\sum_n\frac{-e^{-\beta E_n}}{\mathcal{Z}^2}\delta\mathcal{Z}\langle n|\hat{T}_{ij}|n\rangle \langle n|\hat{T}_{kl}|n\rangle \qquad\qquad\qquad\qquad\qquad\qquad\qquad J_5\nonumber \\
 & {} & -\frac{\beta }{N_0^3V}\sum_n\frac{e^{-\beta E_n}}{\mathcal{Z}}\bigg[\left(\delta\langle n|\right)\hat{T}_{ij}|n\rangle \langle n|\hat{T}_{kl}|n\rangle+\langle n|\hat{T}_{ij}\left(\delta |n\rangle\right)\langle n|\hat{T}_{kl}|n\rangle\nonumber \\
 & {} & \qquad\qquad\qquad\qquad\qquad+\langle n|\hat{T}_{ij} |n\rangle\left(\delta\langle n|\right)\hat{T}_{kl}|n\rangle+\langle n|\hat{T}_{ij} |n\rangle\langle n|\hat{T}_{kl}\left(\delta|n\rangle\right)\bigg]\quad J_6
\end{eqnarray}
Expansion for term $J_4$:
\begin{eqnarray}\label{B11}
J_4 & = & \frac{\beta^2 }{N_0^3V}\sum_n\frac{e^{-\beta E_n}}{\mathcal{Z}}\delta E_n\langle n|\hat{T}_{ij}|n\rangle \langle n|\hat{T}_{kl}|n\rangle\nonumber \\
 & = & \frac{\beta^2}{N_0^3V}\sum_n\frac{e^{-\beta E_n}}{\mathcal{Z}}\sum_{abcd}\sum_{uu'}\sum_{ss'}\Lambda_{abcd}^{(uu')}\langle n|\hat{T}_{ab}^{(u)}\hat{T}_{cd}^{(u')}|n\rangle \langle n|\hat{T}_{ij}^{(s)}|n\rangle \langle n|\hat{T}_{kl}^{(s')}|n\rangle\nonumber \\
 & = & \frac{\beta^2}{N_0^3V}\sum_n\frac{e^{-\beta E_n}}{\mathcal{Z}}\sum_{abcd}\sum_{uu'}\sum_{ss'}\Lambda_{abcd}^{(uu')}\langle n|\hat{T}_{ab}^{(u)}\sum_k|k\rangle\langle k|\hat{T}_{cd}^{(u')}|n\rangle \langle n|\hat{T}_{ij}^{(s)}|n\rangle \langle n|\hat{T}_{kl}^{(s')}|n\rangle\nonumber \\
 & = & \frac{\beta^2}{N_0^3V}\sum_n\frac{e^{-\beta E_n}}{\mathcal{Z}}\sum_{abcd}\sum_{ss'}\Lambda_{abcd}^{(ss')}\langle n^{(s)}|\hat{T}_{ab}^{(s)}|n^{(s)}\rangle\langle n^{(s)}|\hat{T}_{ij}^{(s)}|n^{(s)}\rangle \langle n^{(s')}|\hat{T}_{kl}^{(s')}|n^{(s')}\rangle \langle n^{(s')}|\hat{T}_{cd}^{(s')}|n^{(s')}\rangle\nonumber \\
 & = & \frac{V}{N_0^3}\sum_{abcd}\sum_{ss'}\Lambda_{abcd}^{(ss')}\chi_{abij}^{\rm rel(2)}\chi_{cdkl}^{\rm rel(2)}
\end{eqnarray}
where in the above calculation we insert the identity $\sum_k|k\rangle\langle k|$. Again, the coefficient $\Lambda_{abcd}^{(uu')}$ does not allow $u\neq u'$. We need to pair matrix elements $\langle n|\hat{T}_{ab}^{(u)}|k\rangle$, $\langle k|\hat{T}_{cd}^{(u')}|n\rangle $ with $\langle n|\hat{T}_{ij}^{(s)}|n\rangle$, $\langle n|\hat{T}_{kl}^{(s')}|n\rangle$. Therefore,  we choose $u=s, u'=s'$, or $u=s', u'=s$. Since the matrix elements $\langle n|\hat{T}_{ij}^{(s)}|n\rangle$, $\langle n|\hat{T}_{kl}^{(s')}|n\rangle$ are diagonal, the only choice for quantum number $k$ is $k=n$.\\
Expansion for term $J_5$:
\begin{eqnarray}\label{B12}
J_5 & = & \frac{\beta}{N_0^3V}\sum_n\frac{e^{-\beta E_n}}{\mathcal{Z}^2}\delta \mathcal{Z}\langle n|\hat{T}_{ij}|n\rangle \langle n|\hat{T}_{kl}|n\rangle\nonumber \\
 & = & -\frac{\beta^2}{N_0^3V}\sum_{nl}\frac{e^{-\beta( E_n+E_l)}}{\mathcal{Z}^2}\langle l|\hat{V}|l\rangle \langle n|\hat{T}_{ij}|n\rangle \langle n|\hat{T}_{kl}|n\rangle\nonumber \\
 & = & -\frac{\beta^2}{N_0^3V}\sum_{nl}\frac{e^{-\beta( E_n+E_l)}}{\mathcal{Z}^2}\sum_{abcd}\sum_{uu'}\sum_{ss'}\Lambda_{abcd}^{(uu')}\langle l| \hat{T}_{ab}^{(u)}\hat{T}_{cd}^{(u')}|l\rangle \langle n|\hat{T}_{ij}^{(s)}|n\rangle \langle n|\hat{T}_{kl}^{(s')}|n\rangle\nonumber \\
 & = & -\frac{\beta^2}{N_0^3V}\sum_{nl}\frac{e^{-\beta( E_n+E_l)}}{\mathcal{Z}^2}\sum_{abcd}\sum_{uu'}\sum_{ss'}\Lambda_{abcd}^{(uu')}\langle l| \hat{T}_{ab}^{(u)}\sum_k|k\rangle\langle k|\hat{T}_{cd}^{(u')}|l\rangle \langle n|\hat{T}_{ij}^{(s)}|n\rangle \langle n|\hat{T}_{kl}^{(s')}|n\rangle\nonumber \\
 & = & -\frac{\beta^2}{N_0^3V}\sum_{lmn}\frac{e^{-\beta( E_n+E_l)}}{\mathcal{Z}^2}\sum_{abcd}\sum_{uu'}\sum_{ss'}\Lambda_{abcd}^{(uu')}\langle l| \hat{T}_{ab}^{(u)}|m\rangle \langle m|\hat{T}_{cd}^{(u')}|l\rangle \langle n|\hat{T}_{ij}^{(s)}|n\rangle \langle n|\hat{T}_{kl}^{(s')}|n\rangle\nonumber \\
 & = & -\frac{\beta^2}{N_0^3V}\sum_{ss'}\sum_{l^{(s)}l^{(s')}n^{(s)}n^{(s')}}\frac{e^{-\beta( E_n^{(s)}+E_n^{(s')}+E_l^{(s)}+E_l^{(s')})}}{\mathcal{Z}^{(s)2}\mathcal{Z}^{(s')2}}\sum_{abcd}\Lambda_{abcd}^{(ss')}\nonumber \\
 & {} & \langle l^{(s)}| \hat{T}_{cd}^{(s)}|l^{(s)}\rangle \langle n^{(s)}|\hat{T}_{ij}^{(s)}|n^{(s)}\rangle \langle l^{(s')}|\hat{T}_{ab}^{(s')}|l^{(s')}\rangle \langle n^{(s')}|\hat{T}_{kl}^{(s')}|n^{(s')}\rangle\nonumber \\
 & = & -\frac{V}{N_0^3}\sum_{ss'}\sum_{abcd}\Lambda_{abcd}^{(ss')}\chi_{abij}^{\rm rel(1)}\chi_{cdkl}^{\rm rel(1)}
\end{eqnarray}
where in the above calculations we have inserted the identity $\sum_k|k\rangle\langle k|$. The only choice for quantum number $k$ is $k=l$.\\
Expansion for term $J_6$:
\begin{eqnarray}\label{B13.1}
J_6 & = & -\frac{\beta}{N_0^3V}\sum_{nm}\frac{1}{E_n-E_m}\frac{e^{-\beta E_n}}{\mathcal{Z}}\bigg[
   \langle n|\hat{V}|m\rangle \langle m|\hat{T}_{ij}|n\rangle\langle n|\hat{T}_{kl}|n\rangle
+ \langle n|\hat{T}_{ij}|m\rangle\langle m|\hat{V}|n\rangle\langle n|\hat{T}_{kl}|n\rangle\nonumber \\
 & {} & \qquad\qquad\qquad\qquad\qquad\qquad\quad+ \langle n|\hat{T}_{ij}|n\rangle\langle n|\hat{V}|m\rangle\langle m|\hat{T}_{kl}|n\rangle
+ \langle n|\hat{T}_{ij}|n\rangle\langle n|\hat{T}_{kl}|m\rangle\langle m|\hat{V}|n\rangle
\bigg]\nonumber \\
 & = & 
-\frac{\beta}{N_0^3V}\sum_{nml}\sum_{abcd}\sum_{uu'}\sum_{ss'}\frac{1}{E_n-E_m}\frac{e^{-\beta E_n}}{\mathcal{Z}}\Lambda_{abcd}^{(uu')}\nonumber \\
 & {} & \bigg\{\langle n|\hat{T}_{ab}^{(u)}\sum_l|l\rangle \langle l|\hat{T}_{cd}^{(u')}|m\rangle\langle m|\hat{T}_{ij}^{(s)}|n\rangle\langle n|\hat{T}_{kl}^{(s')}|n\rangle+\langle n|\hat{T}_{ij}^{(s)}|m\rangle\langle m|\hat{T}_{ab}^{(u)}\sum_l|l\rangle \langle l|\hat{T}_{cd}^{(u')}|n\rangle\langle n|\hat{T}_{kl}^{(s')}|n\rangle\nonumber \\
 & {} & +\langle n|\hat{T}_{ij}^{(s)}|n\rangle\langle n|\hat{T}_{ab}^{(u)}\sum_l|l\rangle \langle l|\hat{T}_{cd}^{(u')}|m\rangle\langle m|\hat{T}_{kl}^{(s')}|n\rangle+\langle n|\hat{T}_{ij}^{(s)}|n\rangle\langle n|\hat{T}_{kl}^{(s')}|m\rangle\langle m|\hat{T}_{ab}^{(u)}\sum_l|l\rangle \langle l|\hat{T}_{cd}^{(u')}|n\rangle\bigg\}\nonumber \\
 & = & 
-\frac{\beta}{N_0^3V}\sum_{nml}\sum_{abcd}\sum_{uu'}\sum_{ss'}\frac{1}{E_n-E_m}\frac{e^{-\beta E_n}}{\mathcal{Z}}\Lambda_{abcd}^{(uu')}\nonumber \\
 & {} & \bigg\{\langle n|\hat{T}_{ab}^{(u)}|l\rangle \langle l|\hat{T}_{cd}^{(u')}|m\rangle\langle m|\hat{T}_{ij}^{(s)}|n\rangle\langle n|\hat{T}_{kl}^{(s')}|n\rangle+\langle n|\hat{T}_{ij}^{(s)}|m\rangle\langle m|\hat{T}_{ab}^{(u)}|l\rangle\langle l|\hat{T}_{cd}^{(u')}|n\rangle\langle n|\hat{T}_{kl}^{(s')}|n\rangle\nonumber \\
 & {} & +\langle n|\hat{T}_{ij}^{(s)}|n\rangle\langle n|\hat{T}_{ab}^{(u)}|l\rangle \langle l|\hat{T}_{cd}^{(u')}|m\rangle\langle m|\hat{T}_{kl}^{(s')}|n\rangle+\langle n|\hat{T}_{ij}^{(s)}|n\rangle\langle n|\hat{T}_{kl}^{(s')}|m\rangle\langle m|\hat{T}_{ab}^{(u)}|l\rangle\langle l|T_{cd}^{(u')}|n\rangle\bigg\}\nonumber \\
 & = & 
-\frac{\beta}{N_0^3V}\sum_{nml}\sum_{abcd}\sum_{ss'}\frac{1}{E_n-E_m}\frac{e^{-\beta E_n}}{\mathcal{Z}}\Lambda_{abcd}^{(ss')}\nonumber \\
 & {} & \bigg\{\langle n|\hat{T}_{ab}^{(s')}|l\rangle \langle l|\hat{T}_{cd}^{(s)}|m\rangle\langle m|\hat{T}_{ij}^{(s)}|n\rangle\langle n|\hat{T}_{kl}^{(s')}|n\rangle+\langle n|\hat{T}_{ij}^{(s)}|m\rangle\langle m|\hat{T}_{ab}^{(s)}|l\rangle\langle l|\hat{T}_{cd}^{(s')}|n\rangle\langle n|\hat{T}_{kl}^{(s')}|n\rangle\nonumber \\
 & {} & +\langle n|\hat{T}_{ij}^{(s)}|n\rangle\langle n|\hat{T}_{ab}^{(s)}|l\rangle \langle l|\hat{T}_{cd}^{(s')}|m\rangle\langle m|\hat{T}_{kl}^{(s')}|n\rangle+\langle n|\hat{T}_{ij}^{(s)}|n\rangle\langle n|\hat{T}_{kl}^{(s')}|m\rangle\langle m|\hat{T}_{ab}^{(s')}|l\rangle\langle l|\hat{T}_{cd}^{(s)}|n\rangle\bigg\}
\end{eqnarray}
In the above calculations, we insert the identity $\sum_l|l\rangle\langle l|$. 

There are four summations in the above result. As an example, we discuss the pairing rule for the first summation, $\sum_{nml}\sum_{uu'}\sum_{ss'}\Lambda_{abcd}^{(uu')}\langle n|\hat{T}_{ab}^{(u)}|l\rangle \langle l|\hat{T}_{cd}^{(u')}|m\rangle\langle m|\hat{T}_{ij}^{(s)}|n\rangle\langle n|\hat{T}_{kl}^{(s')}|n\rangle$. The coefficient $\Lambda_{abcd}^{(uu')}$ does not allow $u=u'$. Therefore the matrix elements $\langle n|\hat{T}_{ab}^{(u)}|l\rangle$, $ \langle l|\hat{T}_{cd}^{(u')}|m\rangle$ cannot be paired with each other. We need to pair $\langle n|\hat{T}_{ab}^{(u)}|l\rangle$, $ \langle l|\hat{T}_{cd}^{(u')}|m\rangle$ with $\langle m|\hat{T}_{ij}^{(s)}|n\rangle$, $\langle n|\hat{T}_{kl}^{(s')}|n\rangle$. There are two pairing candidates: first, $\langle n|\hat{T}_{ab}^{(u)}|l\rangle$ is paired with $\langle m|\hat{T}_{ij}^{(s)}|n\rangle$, and $ \langle l|\hat{T}_{cd}^{(u')}|m\rangle$ is paired with $\langle n|\hat{T}_{kl}^{(s')}|n\rangle$; second, $\langle n|\hat{T}_{ab}^{(u)}|l\rangle$ is paired with $\langle n|\hat{T}_{kl}^{(s')}|n\rangle$, and $ \langle l|\hat{T}_{cd}^{(u')}|m\rangle$ is paired with $\langle m|\hat{T}_{ij}^{(s)}|n\rangle$. In the first candidate of pairing matrix elements, we have $u=s, u'=s'$. Since $\langle n|\hat{T}_{kl}^{(s')}|n\rangle$ is diagonal, the matrix element $ \langle l|\hat{T}_{cd}^{(u')}|m\rangle$ which is pair to it must be diagonal as well. Therefore the only choice for quantum number $l$ is $l=m$. In the second candidate, we have $u=s', u'=s$. Again, since $\langle n|\hat{T}_{kl}^{(s')}|n\rangle$ is diagonal, the matrix element $\langle n|\hat{T}_{ab}^{(u)}|l\rangle$ which is paired to it must be diagonal as well. The only choice for quantum number $l$ is $l=n$. With the previous pairing rule, we continue our calculation as follows,  

\begin{eqnarray}\label{B13}
J_6
 & = & 
-\frac{2\beta}{N_0^3V}\sum_{abcd}\sum_{ss'}\sum_{n^{(s)}n^{(s')}m^{(s)}}\frac{1}{E_n^{(s)}-E_m^{(s)}}\frac{e^{-\beta (E_n^{(s)}+E_n^{(s')})}}{\mathcal{Z}^{(s)}\mathcal{Z}^{(s')}}\Lambda_{abcd}^{(ss')}\nonumber \\
 & {} & 
\langle n^{(s')}|\hat{T}_{kl}^{(s')}|n^{(s')}\rangle \langle n^{(s')}|\hat{T}_{ab}^{(s')}|n^{(s')}\rangle\langle m^{(s)}|\hat{T}_{cd}^{(s)}|n^{(s)}\rangle\langle n^{(s)}|\hat{T}_{ij}^{(s)}|m^{(s)}\rangle\nonumber \\
 & + & -\frac{2\beta}{N_0^3V}\sum_{abcd}\sum_{ss'}\sum_{n^{(s)}n^{(s')}m^{(s')}}\frac{1}{E_n^{(s')}-E_m^{(s')}}\frac{e^{-\beta (E_n^{(s)}+E_n^{(s')})}}{\mathcal{Z}^{(s)}\mathcal{Z}^{(s')}}\Lambda_{abcd}^{(ss')}\nonumber \\
 & {} & \langle n^{(s)}|\hat{T}_{ij}^{(s)}|n^{(s)}\rangle\langle n^{(s)}|\hat{T}_{ab}^{(s)}|n^{(s)}\rangle\langle n^{(s')}|\hat{T}_{cd}^{(s')}|m^{(s')}\rangle\langle m^{(s')}|\hat{T}_{kl}^{(s')}|n^{(s')}\rangle\nonumber \\
 & = & -\frac{V}{N_0^3}\sum_{abcd}\sum_{ss'}\Lambda_{abcd}^{(ss')}\left(\chi_{abkl}^{\rm rel(2)}\chi_{cdij}^{\rm res}+\chi_{abij}^{\rm rel(2)}\chi_{cdkl}^{\rm res}\right)
\end{eqnarray}

\subsection{Expansion details for $\chi_{ijkl}^{\rm super \, res}$}
\begin{eqnarray}\label{B14}
\chi_{ijkl}^{\rm super \, res} & = & \frac{2}{\hbar N_0^3V}\sum_{n^*l^*}\frac{e^{-\beta E_n^*}}{\mathcal{Z}^*}\langle l^*|\hat{T}_{ij}|n^*\rangle\langle n^*|\hat{T}_{kl}|l^*\rangle\frac{\omega_{ln}^*}{(i\eta)^2-\omega_{ln}^{*2}}\nonumber \\
 & = & \frac{2}{\hbar N_0^3V}\sum_{nl}\frac{e^{-\beta E_n}}{\mathcal{Z}}\langle l|\hat{T}_{ij}|n\rangle\langle n|\hat{T}_{kl}|l\rangle\frac{\omega_{ln}}{(i\eta)^2-\omega_{ln}^{2}}\nonumber \\
 & + & \frac{2}{\hbar N_0^3V}\sum_{nl}\frac{e^{-\beta E_n}(-\beta \delta E_n)}{\mathcal{Z}}\langle l|\hat{T}_{ij}|n\rangle\langle n|\hat{T}_{kl}|l\rangle\frac{\omega_{ln}}{(i\eta)^2-\omega_{ln}^2}\qquad\qquad\qquad\qquad\qquad\qquad J_7\nonumber \\
 & + & \frac{2}{\hbar N_0^3V}\sum_{nl}\frac{e^{-\beta E_n}(-\delta \mathcal{Z})}{\mathcal{Z}}\langle l|\hat{T}_{ij}|n\rangle\langle n|\hat{T}_{kl}|l\rangle\frac{\omega_{ln}}{(i\eta)^2-\omega_{ln}^2}
\qquad\qquad\qquad\qquad\qquad\qquad\quad\, J_8
\nonumber \\
 & + & \frac{2}{\hbar N_0^3V}\sum_{nl}\frac{e^{-\beta E_n}}{\mathcal{Z}}\langle l|\hat{T}_{ij}|n\rangle\langle n|\hat{T}_{kl}|l\rangle\frac{(i\eta)^2+\omega_{ln}^2}{[(i\eta)^2-\omega_{ln}^2]^2}\delta\omega_{ln}
\qquad\qquad\qquad\qquad\qquad\qquad\quad\, J_9
\nonumber \\
 & + & \frac{2}{\hbar N_0^3V}\sum_{nl}\frac{e^{-\beta E_n}}{\mathcal{Z}}\bigg[\left(\delta\langle l|\right)\hat{T}_{ij}|n\rangle\langle n|\hat{T}_{kl}|l\rangle+\langle l|\hat{T}_{ij}\left(\delta|n\rangle\right)\langle n|\hat{T}_{kl}|l\rangle
\nonumber \\
 & {} & \qquad\qquad\qquad\qquad+\langle l|\hat{T}_{ij}|n\rangle\left(\delta\langle n|\right)\hat{T}_{kl}|l\rangle+\langle l|\hat{T}_{ij}|n\rangle\langle n|\hat{T}_{kl}\left(\delta|l\rangle\right)\bigg]\frac{\omega_{ln}}{(i\eta)^2-\omega_{ln}^2}\qquad\quad J_{10}
\end{eqnarray}
where please note we use the simplified notation $(E_l-E_n)/\hbar=\omega_l-\omega_n=\omega_{ln}$. We denote the change of $\omega_{ln}$ to be $\delta\omega_{ln}=(
\delta E_l-\delta E_n)/\hbar$.\\
Expansion for term $J_7$:
\begin{eqnarray}\label{B15}
J_7 & = & -\frac{2\beta}{\hbar N_0^3V}\sum_{nl}\frac{e^{-\beta E_n}}{\mathcal{Z}}\langle n|\hat{V}|n\rangle\langle l|\hat{T}_{ij}|n\rangle\langle n|\hat{T}_{kl}|l\rangle\frac{\omega_{ln}}{(i\eta)^2-\omega_{ln}^2}\nonumber \\
 & = & -\frac{2\beta}{\hbar N_0^3V}\sum_{nl}\frac{e^{-\beta E_n}}{\mathcal{Z}}\sum_{abcd}\sum_{uu'}\sum_{ss'}\Lambda_{abcd}^{(uu')}\langle n|\hat{T}_{ab}^{(u)}\hat{T}_{cd}^{(u')}|n\rangle\langle l|\hat{T}_{ij}^{(s)}|n\rangle\langle n|\hat{T}_{kl}^{(s')}|l\rangle\frac{\omega_{ln}}{(i\eta)^2-\omega_{ln}^2}\nonumber \\
 & = & -\frac{2\beta}{\hbar N_0^3V}\sum_{nl}\frac{e^{-\beta E_n}}{\mathcal{Z}}\sum_{abcd}\sum_{uu'}\sum_{ss'}\Lambda_{abcd}^{(uu')}\langle n|\hat{T}_{ab}^{(u)}\sum_k|k\rangle\langle k|\hat{T}_{cd}^{(u')}|n\rangle\langle l|\hat{T}_{ij}^{(s)}|n\rangle\langle n|\hat{T}_{kl}^{(s')}|l\rangle\frac{\omega_{ln}}{(i\eta)^2-\omega_{ln}^2}\nonumber \\
 & = & -\frac{4\beta}{\hbar N_0^3V}\sum_{nl}\frac{e^{-\beta E_n}}{\mathcal{Z}}\sum_{abcd}\sum_{ss'}\Lambda_{abcd}^{(ss')}\frac{\omega_{ln}}{(i\eta)^2-\omega_{ln}^2}
{\rm Tr}\,\left[\hat{T}_{cd}^{(s)}|n\rangle\langle l|\hat{T}_{ij}^{(s)}|n\rangle\langle n|\hat{T}_{kl}^{(s')}|l\rangle\langle n|\hat{T}_{ab}^{(s')}\right]\nonumber \\
 & = & 
-\frac{4\beta}{\hbar N_0^3V}\sum_{nl}\frac{e^{-\beta E_n}}{\mathcal{Z}}\sum_{abcd}\sum_{ss'}\Lambda_{abcd}^{(ss')}\frac{\omega_{ln}}{(i\eta)^2-\omega_{ln}^2}\nonumber \\
 & {} & 
\langle n^{(s)}|\hat{T}_{ij}^{(s)}|n^{(s)}\rangle \langle n^{(s)}|\hat{T}_{ab}^{(s)}|n^{(s)}\rangle\langle n^{(s')}|\hat{T}_{cd}^{(s')}|n^{(s')}\rangle\langle n^{(s')}|\hat{T}_{kl}^{(s')}|n^{(s')}\rangle
\langle n^{(s)}|l^{(s)}\rangle\langle n^{(s')}|l^{(s')}\rangle\langle n^{(r)}|l^{(r)}\rangle\nonumber \\
 & = & 0
\end{eqnarray}
In the above calculations we insert the identity $\sum_k|k\rangle\langle k|$. The coefficient $\Lambda_{abcd}^{(uu')}$ does not allow $u= u'$. Therefore we need to pair matrix elements $\langle n|\hat{T}_{ab}^{(u)}|k\rangle$, $\langle k|\hat{T}_{cd}^{(u')}|n\rangle$ with $\langle l|\hat{T}_{ij}^{(s)}|n\rangle$, $\langle n|\hat{T}_{kl}^{(s')}|l\rangle$. There are two candidates: first, $\langle n|\hat{T}_{ab}^{(u)}|k\rangle$ is paired with $\langle l|\hat{T}_{ij}^{(s)}|n\rangle$, and $\langle k|\hat{T}_{cd}^{(u')}|n\rangle$ is paired with $\langle n|\hat{T}_{kl}^{(s')}|l\rangle$; second, $\langle n|\hat{T}_{ab}^{(u)}|k\rangle$ is paired with $\langle n|\hat{T}_{kl}^{(s')}|l\rangle$, and $\langle k|\hat{T}_{cd}^{(u')}|n\rangle$ is paired with $\langle l|\hat{T}_{ij}^{(s)}|n\rangle$. The first candidate allows $u=s, u'=s'$. Since the matrix elements $\langle l|\hat{T}_{ij}^{(s)}|n\rangle$ and $\langle n|\hat{T}_{kl}^{(s')}|l\rangle$ are off-diagonal, the pairing rule requires the quantum number $k$ to be $k=l$. The matrix element product is therefore given by 
\begin{eqnarray}\label{B15.1}
 & {} & \sum_{nl}\frac{e^{-\beta E_n}}{\mathcal{Z}}\sum_{ss'}\langle n|\hat{T}_{ab}^{(s)}|l\rangle\langle l|\hat{T}_{ij}^{(s)}|n\rangle\langle n|\hat{T}_{kl}^{(s')}|l\rangle\langle l|\hat{T}_{cd}^{(s')}|n\rangle\frac{\omega_{ln}}{(i\eta)^2-\omega_{ln}^2}\nonumber \\
 & = & \frac{e^{-\beta \left(E_n^{(s)}+E_n^{(s')}\right)}}{\mathcal{Z}^{(s)}\mathcal{Z}^{(s')}}\sum_{ss'}\sum_{n^{(s)}n^{(s')}l^{(s)}l^{(s')}}\langle n^{(s)}|\hat{T}_{ab}^{(s)}|l^{(s)}\rangle\langle l^{(s)}|\hat{T}_{ij}^{(s)}|n^{(s)}\rangle\langle n^{(s')}|l^{(s')}\rangle \langle l^{(s')}|n^{(s')}\rangle\nonumber \\
 & {} & \langle n^{(s')}|\hat{T}_{kl}^{(s')}|l^{(s')}\rangle\langle l^{(s')}|\hat{T}_{cd}^{(s')}|n^{(s')}\rangle\langle n^{(s)}|l^{(s)}\rangle\langle l^{(s)}|n^{(s)}\rangle\frac{\omega_{ln}^{(s)}+\omega_{ln}^{(s')}}{(i\eta)^2-\left(\omega_{ln}^{(s)}+\omega_{ln}^{(s')}\right)^2}\nonumber \\
 & = & 0 \quad\quad {\rm since}\quad\quad \omega_{ln}^{(s)}=\omega_{ln}^{(s')}=0
\end{eqnarray}
The second candidate requires $u=s', u'=s$ and $k=l=n$, so that all of the matrix elements are diagonal. We also have $\omega_{ln}=0$. Finally, the above result Eq.(\ref{B15}) is zero.

There is an additional qualitative argument which leads us to the same result for term(7) very quickly: suppose $s\neq s'$ and $l\neq n$ in the second step of Eq.(\ref{B15}). The operator for the $s$-th block stress tensor $\hat{T}_{ij}^{(s)}$ changes state from wavefunction $|l\rangle$ to $|n\rangle$, and the $s'$-th block stress tensor $\hat{T}_{kl}^{(s')}$ changes state from $|n\rangle$ to $|l\rangle$. However, since the $s$-th block stress tensor only acts on the $s$-th block wavefunction, and the $s'$-th block stress tensor only acts on the $s'$-th block wavefunction, it is impossible to change state $|l\rangle$ to state $|n\rangle$ and state $|n\rangle$ to state $|l\rangle$ simultaneously by stress tensor operators from two different blocks. The only possibility is that states $|l\rangle$ and $|n\rangle$ are the same state: $|n\rangle=|l\rangle$, which means the wave functions $|l\rangle$ and $|n\rangle$ are not changed by stress tensors $\hat{T}_{ij}^{(s)}$ and $\hat{T}_{kl}^{(s')}$. This argument also leads to the same result of term(7), because the factor $\omega_{ln}=\omega_l-\omega_n=0$ makes term $J_7$ to vanish. \\
Expansion for term $J_8$:
\begin{eqnarray}\label{B16}
J_8 & = & -\frac{2}{\hbar N_0^3V}\sum_{nl}\frac{e^{-\beta E_n}}{\mathcal{Z}^2}\delta\mathcal{Z}\langle l|\hat{T}_{ij}|n\rangle\langle n|\hat{T}_{kl}|l\rangle\frac{\omega_{ln}}{(i\eta)^2-\omega_{ln}^2}\nonumber \\
 & = & \frac{2\beta}{\hbar N_0^3V}\sum_{lmn}\frac{e^{-\beta (E_n+E_m)}}{\mathcal{Z}^2}\langle m|V|m\rangle\langle l|\hat{T}_{ij}|n\rangle\langle n|\hat{T}_{kl}|l\rangle\frac{\omega_{ln}}{(i\eta)^2-\omega_{ln}^2}\nonumber \\
 & = & \frac{2\beta}{\hbar N_0^3V}\sum_{lmn}\frac{e^{-\beta (E_n+E_m)}}{\mathcal{Z}^2}\sum_{abcd}\sum_{uu'}\sum_{ss'}\Lambda_{abcd}^{(uu')}\frac{\omega_{ln}}{(i\eta)^2-\omega_{ln}^2}\langle m|\hat{T}_{ab}^{(u)}\hat{T}_{cd}^{(u')}|m\rangle\langle l|\hat{T}_{ij}^{(s)}|n\rangle\langle n|\hat{T}_{kl}^{(s')}|l\rangle\nonumber \\
 & = & \frac{2\beta}{\hbar N_0^3V}\sum_{lmn}\frac{e^{-\beta (E_n+E_m)}}{\mathcal{Z}^2}\sum_{abcd}\sum_{uu'}\sum_{ss'}\Lambda_{abcd}^{(uu')}\frac{\omega_{ln}}{(i\eta)^2-\omega_{ln}^2}\langle m|\hat{T}_{ab}^{(u)}\sum_k|k\rangle \langle k|\hat{T}_{cd}^{(u')}|m\rangle\langle l|\hat{T}_{ij}^{(s)}|n\rangle\langle n|\hat{T}_{kl}^{(s')}|l\rangle\nonumber \\
 & = & \frac{2\beta}{\hbar N_0^3V}\sum_{lmn}\frac{e^{-\beta (E_n+E_m)}}{\mathcal{Z}^2}\sum_{abcd}\sum_{uu'}\sum_{ss'}\Lambda_{abcd}^{(uu')}\frac{\omega_{ln}}{(i\eta)^2-\omega_{ln}^2}{\rm Tr}\,\left[\hat{T}_{cd}^{(u')}|m\rangle\langle l|\hat{T}_{ij}^{(s)}|n\rangle\langle n|\hat{T}_{kl}^{(s')}|l\rangle\langle m|\hat{T}_{ab}^{(u)}\right]\nonumber \\
 & = & \frac{4\beta}{\hbar N_0^3V}\sum_{lmn}\frac{e^{-\beta (E_n+E_m)}}{\mathcal{Z}^2}\sum_{abcd}\sum_{uu'}\sum_{ss'}\Lambda_{abcd}^{(uu')}\frac{\omega_{ln}}{(i\eta)^2-\omega_{ln}^2}\nonumber \\
 & {} & \langle m^{(s)}|\hat{T}_{cd}^{(s)}|m^{(s)}\rangle\langle l^{(s)}|\hat{T}_{ij}^{(s)}|n^{(s)}\rangle \langle n^{(s')}|\hat{T}_{kl}^{(s')}|n^{(s')}\rangle \langle m^{(s')}|\hat{T}_{ab}^{(s')}|m^{(s')}\rangle
\langle n^{(s)}|l^{(s)}\rangle\langle n^{(s')}|l^{(s')}\rangle\langle n^{(r)}|l^{(r)}\rangle\nonumber \\
 & = & 0
\end{eqnarray}
Again we insert the identity $\sum_k|k\rangle\langle k|$ in the fourth step of the above calculations. We use the same argument in term $J_7$ calculatins: suppose $s\neq s'$ and $l\neq n$ in the third step of Eq.(\ref{B16}). The operator for the $s$-th block stress tensor $\hat{T}_{ij}^{(s)}$ changes state from wavefunction $|l\rangle$ to $|n\rangle$, and the $s'$-th block stress tensor $\hat{T}_{kl}^{(s')}$ changes state from $|n\rangle$ to $|l\rangle$. However, since the $s$-th block stress tensor only acts on the $s$-th block wavefunction, and the $s'$-th block stress tensor only acts on the $s'$-th block wavefunction, it is impossible to change state $|l\rangle$ to state $|n\rangle$ and state $|n\rangle$ to state $|l\rangle$ simultaneously, except for the only possibility that $|l\rangle$ and $|n\rangle$ are the same. Finally, the factor $\omega_{ln}=\omega_l-\omega_n=0$ makes term $J_8$ to vanish. \\

Expansion for term $J_9$:
\begin{eqnarray}\label{B17}
J_9 & = & \frac{2}{\hbar N_0^3V}\sum_{nl}\frac{e^{-\beta E_n}}{\mathcal{Z}}\langle l|\hat{T}_{ij}|n\rangle\langle n|\hat{T}_{kl}|l\rangle\frac{(i\eta)^2+\omega_{ln}^2}{[(i\eta)^2-\omega_{ln}^2]^2}\delta\omega_{ln}\nonumber \\
 & = & \frac{2}{\hbar^2 N_0^3V}\sum_{nl}\frac{e^{-\beta E_n}}{\mathcal{Z}}\langle l|\hat{T}_{ij}|n\rangle\langle n|\hat{T}_{kl}|l\rangle\frac{(i\eta)^2+\omega_{ln}^2}{[(i\eta)^2-\omega_{ln}^2]^2}\left(\langle l|\hat{V}|l\rangle -\langle n|\hat{V}|n\rangle\right)\nonumber \\
 & = & \frac{2}{\hbar^2 N_0^3V}\sum_{nl}\frac{e^{-\beta E_n}}{\mathcal{Z}}\sum_{abcd}\sum_{uu'}\sum_{ss'}\Lambda_{abcd}^{(uu')}\frac{(i\eta)^2+\omega_{ln}^2}{[(i\eta)^2-\omega_{ln}^2]^2}\langle l|\hat{T}^{(s)}_{ij}|n\rangle\langle n|\hat{T}^{(s')}_{kl}|l\rangle\left(\langle l|\hat{T}_{ab}^{(u)}\hat{T}_{cd}^{(u')}|l\rangle -\langle n|\hat{T}_{ab}^{(u)}\hat{T}_{cd}^{(u')}|n\rangle\right)\nonumber \\
 & = & \frac{2}{\hbar^2 N_0^3V}\sum_{nl}\frac{e^{-\beta E_n}}{\mathcal{Z}}\sum_{abcd}\sum_{uu'}\sum_{ss'}\Lambda_{abcd}^{(uu')}\frac{(i\eta)^2+\omega_{ln}^2}{[(i\eta)^2-\omega_{ln}^2]^2}\nonumber \\
 & {} & \left[{\rm Tr}\,\left(\hat{T}_{cd}^{(u')}|l\rangle\langle l|\hat{T}_{ij}^{(s)}|n\rangle\langle n|\hat{T}_{kl}^{(s')}|l\rangle \langle l|\hat{T}_{ab}^{(u')}\right)-{\rm Tr}\,\left(\hat{T}_{cd}^{(u')}|n\rangle\langle l|\hat{T}_{ij}^{(s)}|n\rangle\langle n|\hat{T}_{kl}^{(s')}|l\rangle\langle n|\hat{T}_{ab}^{(u)}\right)\right]\nonumber \\
 & = & \frac{1}{\hbar^2 N_0^3V}\sum_{nl}\frac{e^{-\beta E_n}}{\mathcal{Z}}\sum_{abcd}\sum_{uu'}\sum_{ss'}\Lambda_{abcd}^{(uu')}\frac{(i\eta)^2+\omega_{ln}^2}{[(i\eta)^2-\omega_{ln}^2]^2}
\langle n^{(s)}|l^{(s)}\rangle\langle n^{(s')}|l^{(s')}\rangle\langle n^{(r)}|l^{(r)}\rangle\nonumber \\
 & {} & \bigg[\langle n^{(s)}|\hat{T}_{ab}^{(s)}|n^{(s)}\rangle\langle n^{(s)}|\hat{T}_{ij}^{(s)}|n^{(s)}\rangle\langle n^{(s')}|\hat{T}_{cd}^{(s')}|n^{(s')}\rangle\langle n^{(s')}|\hat{T}_{kl}^{(s')}|n^{(s')}\rangle\nonumber \\
 & {} & -\langle n^{(s)}|\hat{T}_{ab}^{(s)}|n^{(s)}\rangle\langle n^{(s)}|\hat{T}_{ij}^{(s)}|n^{(s)}\rangle\langle n^{(s')}|\hat{T}_{cd}^{(s')}|n^{(s')}\rangle\langle n^{(s')}|\hat{T}_{kl}^{(s')}|n^{(s')}\rangle\bigg]\nonumber \\
 & = & 0
\end{eqnarray}
To obtain the above vanishing result of $J_9$ we use the same argument in calculating $J_7$ and $J_8$.

Expansion for term $J_{10}$:
\begin{eqnarray}\label{B18.1}
J_{10} & = & \frac{2}{\hbar N_0^3V}\sum_{nl}\frac{e^{-\beta E_n}}{\mathcal{Z}}\bigg[\left(\delta\langle l|\right)\hat{T}_{ij}|n\rangle\langle n|\hat{T}_{kl}|l\rangle+\langle l|\hat{T}_{ij}\left(\delta|n\rangle\right)\langle n|\hat{T}_{kl}|l\rangle\nonumber \\
 & {} & \qquad\qquad\qquad
+\langle l|\hat{T}_{ij}|n\rangle\left(\delta\langle n|\right)\hat{T}_{kl}|l\rangle+\langle l|\hat{T}_{ij}|n\rangle\langle n|\hat{T}_{kl}\left(\delta|l\rangle\right)\bigg]\frac{\omega_{ln}}{(i\eta)^2-\omega_{ln}^2}\nonumber \\
 & = & \frac{2}{\hbar N_0^3V}\sum_{nl}\sum_{abcd}\sum_{uu'}\sum_{ss'}\frac{e^{-\beta E_n}}{\mathcal{Z}}\Lambda_{abcd}^{(uu')}\frac{\omega_{ln}}{(i\eta)^2-\omega_{ln}^2}\sum_m\nonumber \\
 & {} & \bigg\{\frac{1}{E_l-E_m}\left(\langle l|\hat{T}_{ab}^{(u)}\hat{T}_{cd}^{(u')}|m\rangle\langle m|\hat{T}_{ij}^{(s)}|n\rangle\langle n|\hat{T}_{kl}^{(s')}|l\rangle+\langle l|\hat{T}_{ij}^{(s)}|n\rangle\langle n|\hat{T}_{kl}^{(s')}|m\rangle\langle m|\hat{T}_{ab}^{(u)}\hat{T}_{cd}^{(u')}|l\rangle\right)\nonumber \\
 & {} & +\frac{1}{E_n-E_m}\left(\langle l|\hat{T}_{ij}^{(s)}|m\rangle\langle m|\hat{T}_{ab}^{(u)}\hat{T}_{cd}^{(u')}|n\rangle\langle n|\hat{T}_{kl}^{(s')}|l\rangle+\langle l|\hat{T}_{ij}^{(s)}|n\rangle \langle n|\hat{T}_{ab}^{(u)}\hat{T}_{cd}^{(u')}|m\rangle\langle m|\hat{T}_{kl}^{(s')}|l\rangle\right)\bigg\}\nonumber \\
 & = & \frac{2}{\hbar N_0^3V}\sum_{nl}\sum_{abcd}\sum_{uu'}\sum_{ss'}\frac{e^{-\beta E_n}}{\mathcal{Z}}\Lambda_{abcd}^{(uu')}\frac{\omega_{ln}}{(i\eta)^2-\omega_{ln}^2}\sum_m\sum_k\nonumber \\
 & \bigg\{ & \frac{1}{E_l-E_m}\left(\langle l|\hat{T}_{ab}^{(u)}|k\rangle\langle k|\hat{T}_{cd}^{(u')}|m\rangle\langle m|\hat{T}_{ij}^{(s)}|n\rangle\langle n|\hat{T}_{kl}^{(s')}|l\rangle+\langle l|\hat{T}_{ij}^{(s)}|n\rangle\langle n|\hat{T}_{kl}^{(s')}|m\rangle\langle m|\hat{T}_{ab}^{(u)}|k\rangle\langle k|\hat{T}_{cd}^{(u')}|l\rangle\right)\nonumber \\
 & + & \frac{1}{E_n-E_m}\left(\langle l|\hat{T}_{ij}^{(s)}|m\rangle\langle m|\hat{T}_{ab}^{(u)}|k\rangle\langle k|\hat{T}_{cd}^{(u')}|n\rangle\langle n|\hat{T}_{kl}^{(s')}|l\rangle+\langle l|\hat{T}_{ij}^{(s)}|n\rangle \langle n|\hat{T}_{ab}^{(u)}|k\rangle\langle k|\hat{T}_{cd}^{(u')}|m\rangle\langle m|\hat{T}_{kl}^{(s')}|l\rangle\right)\bigg\}
\end{eqnarray}
where in the third step of the above calculation we insert the identity $\sum_k|k\rangle\langle k|$. Because of the coefficient $\Lambda_{abcd}^{(uu')}$, we have $u\neq u'$. In the final result of the above Eq.(\ref{B18.1}) we get 4 summations. Let us discuss the first  summation for example. 
\begin{eqnarray}\label{B18.2}
 & {} & \frac{2}{\hbar N_0^3V}\sum_{mnkl}\sum_{abcd}\sum_{uu'}\sum_{ss'}\frac{e^{-\beta E_n}}{\mathcal{Z}}\Lambda_{abcd}^{(uu')}\frac{\omega_{ln}}{(i\eta)^2-\omega_{ln}^2}\frac{1}{E_l-E_m}\langle l|\hat{T}_{ab}^{(u)}|k\rangle\langle k|\hat{T}_{cd}^{(u')}|m\rangle\langle m|\hat{T}_{ij}^{(s)}|n\rangle\langle n|\hat{T}_{kl}^{(s')}|l\rangle
\end{eqnarray}
The pairing rule for the other three summations are the same. In the summation, Eq.(\ref{B18.2}), the matrix elements $\langle l|\hat{T}_{ab}^{(u)}|k\rangle$, $\langle k|\hat{T}_{cd}^{(u')}|m\rangle$ must be paired with $\langle m|\hat{T}_{ij}^{(s)}|n\rangle$, $\langle n|\hat{T}_{kl}^{(s')}|l\rangle$. We get two candidates of pairing: first, $\langle l|\hat{T}_{ab}^{(u)}|k\rangle$ is paired with $\langle m|\hat{T}_{ij}^{(s)}|n\rangle$, and $\langle k|\hat{T}_{cd}^{(u')}|m\rangle$ is paired with $\langle n|\hat{T}_{kl}^{(s')}|l\rangle$; second, $\langle l|\hat{T}_{ab}^{(u)}|k\rangle$ is paired with $\langle n|\hat{T}_{kl}^{(s')}|l\rangle$, and $\langle k|\hat{T}_{cd}^{(u')}|m\rangle$ is paired with $\langle m|\hat{T}_{ij}^{(s)}|n\rangle$.

In the first candidate, we have $u=s, u'=s'$. According to the factor ${\omega_{ln}}/({(i\eta)^2-\omega_{ln}^2})$ which requires $l\neq n$, the matrix element $\langle n|\hat{T}_{kl}^{(s')}|l\rangle$ must be off-diagonal. Therefore, the matrix element $\langle k|\hat{T}_{cd}^{(u'=s')}|m\rangle$ which is paired to it must be off-diagonal as well. The wavefunctions of $|k\rangle=\prod_{r=1}^{N_0^3}|k^{(r)}\rangle$ and $|m\rangle=\prod_{r=1}^{N_0^3}|m^{(r)}\rangle$ are required to be $|k^{(s')}\rangle=|l^{(s')}\rangle$, $|m^{(s')}\rangle=|n^{(s')}\rangle$, and $\prod_{r\neq s'}|k^{(r)}\rangle=\prod_{r\neq s'}|m^{(r)}\rangle$, $\prod_{r\neq s'}|n^{(r)}\rangle=\prod_{r\neq s'}|l^{(r)}\rangle$. Therefore in the first candidate case, the first term in Eq.(\ref{B18.2}) is simplified as 
\begin{eqnarray}\label{B18.3}
 & {} & \frac{2}{\hbar N_0^3V}\sum_{mnkl}\sum_{abcd}\sum_{ss'}\frac{e^{-\beta E_n}}{\mathcal{Z}}\Lambda_{abcd}^{(ss')}\frac{\omega_{ln}}{(i\eta)^2-\omega_{ln}^2}\frac{1}{E_l-E_m}\langle l|\hat{T}_{ab}^{(s)}|k\rangle\langle m|\hat{T}_{ij}^{(s)}|n\rangle\langle k|\hat{T}_{cd}^{(s')}|m\rangle\langle n|\hat{T}_{kl}^{(s')}|l\rangle\nonumber \\
 & = & \frac{2}{\hbar N_0^3V}\sum_{abcd}\sum_{ss'}\sum_{l^{(s')}m^{(s)}n^{(s)}n^{(s')}}\frac{e^{-\beta (E_n^{(s)}+E_n^{(s')})}}{\mathcal{Z}^{(s)}\mathcal{Z}^{(s')}}\Lambda_{abcd}^{(ss')}\frac{\omega_{ln}^{(s')}}{(i\eta)^2-\omega_{ln}^{(s')2}}\nonumber \\
 & {} & 
\frac{\langle n^{(s)}|\hat{T}_{ab}^{(s)}|m^{(s)}\rangle\langle m^{(s)}|\hat{T}_{ij}^{(s)}|n^{(s)}\rangle\langle n^{(s')}|\hat{T}_{kl}^{(s')}|l^{(s')}\rangle\langle l^{(s')}|\hat{T}_{cd}^{(s')}|n^{(s')}\rangle}{(E_n^{(s)}-E_m^{(s)})+(E_l^{(s')}-E_n^{(s')})}\nonumber \\
 & = & \frac{2}{\hbar N_0^3V}\sum_{abcd}\sum_{ss'}\sum_{l^{(s')}m^{(s)}n^{(s)}n^{(s')}}\frac{e^{-\beta (E_n^{(s)}+E_n^{(s')})}}{\mathcal{Z}^{(s)}\mathcal{Z}^{(s')}}\Lambda_{abcd}^{(ss')}\frac{\omega_{ln}^{(s')}}{(i\eta)^2-\omega_{ln}^{(s')2}}\nonumber \\
 & {} & 
\frac{\langle n^{(s)}|\hat{T}_{cd}^{(s)}|m^{(s)}\rangle\langle m^{(s)}|\hat{T}_{ij}^{(s)}|n^{(s)}\rangle\langle n^{(s')}|\hat{T}_{kl}^{(s')}|l^{(s')}\rangle\langle l^{(s')}|\hat{T}_{ab}^{(s')}|n^{(s')}\rangle}{(E_n^{(s)}-E_m^{(s)})+(E_l^{(s')}-E_n^{(s')})}
\end{eqnarray}
where in the last step, we exchange the indices $(ab)$ and $(cd)$ in the stress tensors $\hat{T}_{ab}^{(s)}$ and $\hat{T}_{cd}^{(s')}$. The exchange of indices is correct, because the coefficient $\Lambda_{abcd}^{(ss')}$ have the symmetry property: $\Lambda_{abcd}^{(ss')}=\Lambda_{cdab}^{(ss')}$.

Next we consider the second candidate, with $u=s', u'=s$. Actually the second candidate equals to first candidate, because with the exchange of indices $(ab)$, $(cd)$ and $(s)$, $(s')$, the coefficient $\sum_{ss'}\Lambda_{abcd}^{(ss')}$ keeps invariant: $\sum_{ss'}\Lambda_{ijkl}^{(ss')}=\sum_{ss'}\Lambda_{ijkl}^{(s's)}$, and the stress tensor operators commute: $\left[\hat{T}_{ab}^{(u)},\hat{T}_{cd}^{(u')}\right]_{u\neq u'}=0$.

Repeat the same process for the other three summations in Eq.(\ref{B18.3}), we procede our calculation as follows,
\begin{eqnarray}\label{B18}
J_{10} & = & \frac{4}{\hbar N_0^3V}\sum_{abcd}\sum_{ss'}\sum_{l^{(s')}m^{(s)}n^{(s)}n^{(s')}}\frac{e^{-\beta (E_n^{(s)}+E_n^{(s')})}}{\mathcal{Z}^{(s)}\mathcal{Z}^{(s')}}\Lambda_{abcd}^{(ss')}\frac{\omega_{ln}^{(s')}}{(i\eta)^2-\omega_{ln}^{(s')2}}\nonumber \\
 & {} & 
\frac{\langle n^{(s)}|\hat{T}_{cd}^{(s)}|m^{(s)}\rangle\langle m^{(s)}|\hat{T}_{ij}^{(s)}|n^{(s)}\rangle\langle n^{(s')}|\hat{T}_{kl}^{(s')}|l^{(s')}\rangle\langle l^{(s')}|\hat{T}_{ab}^{(s')}|n^{(s')}\rangle}{(E_n^{(s)}-E_m^{(s)})+(E_l^{(s')}-E_n^{(s')})}\nonumber \\
 & + & \frac{4}{\hbar N_0^3V}\sum_{abcd}\sum_{ss'}\sum_{l^{(s)}m^{(s')}n^{(s)}n^{(s')}}\frac{e^{-\beta (E_n^{(s)}+E_n^{(s')})}}{\mathcal{Z}^{(s)}\mathcal{Z}^{(s')}}\Lambda_{abcd}^{(ss')}\frac{\omega_{ln}^{(s)}}{(i\eta)^2-\omega_{ln}^{(s)2}}\nonumber \\
 & {} & 
\frac{\langle n^{(s)}|\hat{T}_{cd}^{(s)}|l^{(s)}\rangle\langle l^{(s)}|\hat{T}_{ij}^{(s)}|n^{(s)}\rangle\langle n^{(s')}|\hat{T}_{kl}^{(s')}|m^{(s')}\rangle\langle m^{(s')}|\hat{T}_{ab}^{(s')}|n^{(s')}\rangle}{(E_n^{(s')}-E_m^{(s')})+(E_l^{(s)}-E_n^{(s)})}\nonumber \\
 & + & \frac{4}{\hbar N_0^3V}\sum_{abcd}\sum_{ss'}\sum_{l^{(s')}m^{(s)}n^{(s)}n^{(s')}}\frac{e^{-\beta (E_n^{(s)}+E_n^{(s')})}}{\mathcal{Z}^{(s)}\mathcal{Z}^{(s')}}\Lambda_{abcd}^{(ss')}\frac{\omega_{ln}^{(s')}}{(i\eta)^2-\omega_{ln}^{(s')2}}\nonumber \\
 & {} & 
\frac{\langle n^{(s)}|\hat{T}_{cd}^{(s)}|m^{(s)}\rangle\langle m^{(s)}|\hat{T}_{ij}^{(s)}|n^{(s)}\rangle\langle n^{(s')}|\hat{T}_{kl}^{(s')}|l^{(s')}\rangle\langle l^{(s')}|\hat{T}_{ab}^{(s')}|n^{(s')}\rangle}{(E_n^{(s)}-E_m^{(s)})-(E_l^{(s')}-E_n^{(s')})}\nonumber \\
 & + & \frac{4}{\hbar N_0^3V}\sum_{abcd}\sum_{ss'}\sum_{l^{(s)}m^{(s')}n^{(s)}n^{(s')}}\frac{e^{-\beta (E_n^{(s)}+E_n^{(s')})}}{\mathcal{Z}^{(s)}\mathcal{Z}^{(s')}}\Lambda_{abcd}^{(ss')}\frac{\omega_{ln}^{(s)}}{(i\eta)^2-\omega_{ln}^{(s)2}}\nonumber \\
 & {} & 
\frac{\langle n^{(s)}|\hat{T}_{cd}^{(s)}|l^{(s)}\rangle\langle l^{(s)}|\hat{T}_{ij}^{(s)}|n^{(s)}\rangle\langle n^{(s')}|\hat{T}_{kl}^{(s')}|m^{(s')}\rangle\langle m^{(s')}|\hat{T}_{ab}^{(s')}|n^{(s')}\rangle}{(E_n^{(s')}-E_m^{(s')})-(E_l^{(s)}-E_n^{(s)})}
\end{eqnarray}

There are 4 terms above. The third and fourth terms are similar with the first and second terms. Therefore let us focus on the calculations of the first and second terms. To calculate the first term, we exchange the indices $l,m$ and $s,s'$ in it. Because $\sum_{ss'}\Lambda_{abcd}^{(ss')}=\sum_{ss'}\Lambda_{cdab}^{(s's)}$, the first term keeps invariant with the exchange of indices $l$, $m$ and $s$, $s'$:
\begin{eqnarray}\label{B19}
 & {} & \frac{4}{\hbar N_0^3V}\sum_{abcd}\sum_{ss'}\sum_{l^{(s')}m^{(s)}n^{(s)}n^{(s')}}\frac{e^{-\beta (E_n^{(s)}+E_n^{(s')})}}{\mathcal{Z}^{(s)}\mathcal{Z}^{(s')}}\Lambda_{abcd}^{(ss')}\frac{\omega_{ln}^{(s')}}{(i\eta)^2-\omega_{ln}^{(s')2}}\nonumber \\
 & {} & 
\frac{\langle n^{(s)}|\hat{T}_{cd}^{(s)}|m^{(s)}\rangle\langle m^{(s)}|\hat{T}_{ij}^{(s)}|n^{(s)}\rangle\langle n^{(s')}|\hat{T}_{kl}^{(s')}|l^{(s')}\rangle\langle l^{(s')}|\hat{T}_{ab}^{(s')}|n^{(s')}\rangle}{(E_n^{(s)}-E_m^{(s)})+(E_l^{(s')}-E_n^{(s')})}\nonumber \\
 & = &
 \frac{4}{\hbar N_0^3V}\sum_{abcd}\sum_{ss'}\sum_{l^{(s')}m^{(s)}n^{(s)}n^{(s')}}\frac{e^{-\beta (E_n^{(s)}+E_n^{(s')})}}{\mathcal{Z}^{(s)}\mathcal{Z}^{(s')}}\Lambda_{abcd}^{(ss')}\frac{\omega_{mn}^{(s)}}{(i\eta)^2-\omega_{mn}^{(s)2}}\nonumber \\
 & {} & 
\frac{\langle n^{(s')}|\hat{T}_{cd}^{(s')}|l^{(s')}\rangle\langle l^{(s')}|\hat{T}_{ij}^{(s')}|n^{(s')}\rangle\langle n^{(s)}|\hat{T}_{kl}^{(s)}|m^{(s)}\rangle\langle m^{(s)}|\hat{T}_{ab}^{(s)}|n^{(s)}\rangle}{(E_n^{(s')}-E_l^{(s')})+(E_m^{(s)}-E_n^{(s)})}\nonumber \\
 & = &
 \frac{4}{\hbar N_0^3V}\sum_{abcd}\sum_{ss'}\sum_{l^{(s')}m^{(s)}n^{(s)}n^{(s')}}\frac{e^{-\beta (E_n^{(s)}+E_n^{(s')})}}{\mathcal{Z}^{(s)}\mathcal{Z}^{(s')}}\Lambda_{abcd}^{(ss')}\frac{\omega_{nm}^{(s)}}{(i\eta)^2-\omega_{nm}^{(s)2}}\nonumber \\
 & {} & 
\frac{\langle n^{(s')}|\hat{T}_{cd}^{(s')}|l^{(s')}\rangle\langle l^{(s')}|\hat{T}_{ij}^{(s')}|n^{(s')}\rangle\langle n^{(s)}|\hat{T}_{kl}^{(s)}|m^{(s)}\rangle\langle m^{(s)}|\hat{T}_{ab}^{(s)}|n^{(s)}\rangle}{(E_l^{(s')}-E_n^{(s')})+(E_n^{(s)}-E_m^{(s)})}
\end{eqnarray}
Use the identity 
\begin{eqnarray}\label{B23}
\left(\frac{1}{i\eta-\omega_{ln}^{(s')}}-\frac{1}{i\eta+\omega_{nm}^{(s)}}\right)\frac{1}{\omega_{ln}^{(s')}+\omega_{nm}^{(s)}}=\frac{1}{i\eta-\omega_{ln}^{(s')}}\frac{1}{i\eta+\omega_{nm}^{(s)}}
\end{eqnarray}
Finally the first term equals to 
\begin{eqnarray}\label{B24}
 & {} & \frac{1}{\hbar^2 N_0^3V}\sum_{abcd}\sum_{ss'}\sum_{l^{(s')}m^{(s)}n^{(s)}n^{(s')}}\frac{e^{-\beta (E_n^{(s)}+E_n^{(s')})}}{\mathcal{Z}^{(s)}\mathcal{Z}^{(s')}}\Lambda_{abcd}^{(ss')}\nonumber \\
 & {} & 
{\langle n^{(s)}|\hat{T}_{cd}^{(s)}|m^{(s)}\rangle\langle m^{(s)}|\hat{T}_{ij}^{(s)}|n^{(s)}\rangle\langle n^{(s')}|\hat{T}_{kl}^{(s')}|l^{(s')}\rangle\langle l^{(s')}|\hat{T}_{ab}^{(s')}|n^{(s')}\rangle}\nonumber \\
 & {} & \left[\frac{1}{(i\eta-\omega_{ln}^{(s')})}\frac{1}{(i\eta+\omega_{nm}^{(s)})}+
\frac{1}{(i\eta+\omega_{ln}^{(s')})}\frac{1}{(i\eta-\omega_{nm}^{(s)})}
\right]
\end{eqnarray}
Similarly the second term is 
\begin{eqnarray}\label{B25}
 & {} & \frac{1}{\hbar^2 N_0^3V}\sum_{abcd}\sum_{ss'}\sum_{l^{(s')}m^{(s)}n^{(s)}n^{(s')}}\frac{e^{-\beta (E_n^{(s)}+E_n^{(s')})}}{\mathcal{Z}^{(s)}\mathcal{Z}^{(s')}}\Lambda_{abcd}^{(ss')}\nonumber \\
 & {} & 
{\langle n^{(s)}|\hat{T}_{cd}^{(s)}|l^{(s)}\rangle\langle l^{(s)}|\hat{T}_{ij}^{(s)}|n^{(s)}\rangle\langle n^{(s')}|\hat{T}_{kl}^{(s')}|m^{(s')}\rangle\langle m^{(s')}|\hat{T}_{ab}^{(s')}|n^{(s')}\rangle}\nonumber \\
 & {} & \left[\frac{1}{(i\eta-\omega_{ln}^{(s)})}\frac{1}{(i\eta+\omega_{nm}^{(s')})}+
\frac{1}{(i\eta+\omega_{ln}^{(s)})}\frac{1}{(i\eta-\omega_{nm}^{(s')})}
\right]
\end{eqnarray}
The third term, 
\begin{eqnarray}\label{B26}
 & {} & -\frac{1}{\hbar^2 N_0^3V}\sum_{abcd}\sum_{ss'}\sum_{l^{(s')}m^{(s)}n^{(s)}n^{(s')}}\frac{e^{-\beta (E_n^{(s)}+E_n^{(s')})}}{\mathcal{Z}^{(s)}\mathcal{Z}^{(s')}}\Lambda_{abcd}^{(ss')}\nonumber \\
 & {} & 
{\langle n^{(s)}|\hat{T}_{cd}^{(s)}|m^{(s)}\rangle\langle m^{(s)}|\hat{T}_{ij}^{(s)}|n^{(s)}\rangle\langle n^{(s')}|\hat{T}_{kl}^{(s')}|l^{(s')}\rangle\langle l^{(s')}|\hat{T}_{ab}^{(s')}|n^{(s')}\rangle}\nonumber \\
 & {} & \left[\frac{1}{(i\eta+\omega_{ln}^{(s')})}\frac{1}{(i\eta+\omega_{nm}^{(s)})}+
\frac{1}{(i\eta-\omega_{ln}^{(s')})}\frac{1}{(i\eta-\omega_{nm}^{(s)})}
\right]
\end{eqnarray}
The fourth term
\begin{eqnarray}\label{B27}
 & {} & -\frac{1}{\hbar^2 N_0^3V}\sum_{abcd}\sum_{ss'}\sum_{l^{(s')}m^{(s)}n^{(s)}n^{(s')}}\frac{e^{-\beta (E_n^{(s)}+E_n^{(s')})}}{\mathcal{Z}^{(s)}\mathcal{Z}^{(s')}}\Lambda_{abcd}^{(ss')}\nonumber \\
 & {} & 
{\langle n^{(s)}|\hat{T}_{cd}^{(s)}|l^{(s)}\rangle\langle l^{(s)}|\hat{T}_{ij}^{(s)}|n^{(s)}\rangle\langle n^{(s')}|\hat{T}_{kl}^{(s')}|m^{(s')}\rangle\langle m^{(s')}|\hat{T}_{ab}^{(s')}|n^{(s')}\rangle}\nonumber \\
 & {} & \left[\frac{1}{(i\eta+\omega_{ln}^{(s)})}\frac{1}{(i\eta+\omega_{nm}^{(s')})}+
\frac{1}{(i\eta-\omega_{ln}^{(s)})}\frac{1}{(i\eta-\omega_{nm}^{(s')})}
\right]
\end{eqnarray}
Sum the above 4 terms up we finally obtain the result
\begin{eqnarray}\label{B28}
 & {} & J_{10}=\frac{V}{N_0^3}\sum_{abcd}\sum_{ss'}\Lambda_{abcd}^{(ss')}\chi^{\rm res}_{cdij}(i\eta)\chi^{\rm res}_{klab}(i\eta)
\end{eqnarray}

\section{Discussions about the Coefficient of Non-elastic Stress-Stress Interaction $\Lambda_{ijkl}(\vec x_s-\vec x_s')$}
In 1976, Joffrin and Levelut\cite{Joffrin1976} gave a detailed derivation on the coefficient of non-elastic stress-stress interaction, $\Lambda_{ijkl}(\vec x_s-\vec x_s')$. The general Hamiltonian of amorphous solid is the summation of long wavelength phonon Hamiltonian, phonon strain field-stress tensor coupling and the non-elastic part of glass Hamiltonian:
\begin{eqnarray}\label{A2}
\hat{H}= \sum_{\vec k\mu}\left(\frac{| {p}_{\mu}(\vec k)|^2}{2m}+\frac{1}{2}m\omega^2_{\vec k\mu}| {u}_{\mu}(\vec k)|^2\right)+\sum_s\sum_{ij}e_{ij}^{(s)}\hat{T}_{ij}^{(s)}+\hat{H}_{0}^{\rm non}
\end{eqnarray}
where $\mu$ is phonon polarization, i.e., longitudinal and transverse phonon modes; $\vec k$ is the phonon wave number and $m$ is the mass of elementary glass block, $ {p}_{\mu}(\vec k)$ and $ {u}_{\mu}(\vec k)$ are the momentum and displacement operators of elementary glass block, respectively. Phonon strain field $e_{ij}^{(s)}$ is defined as $e_{ij}^{(s)}=\frac{1}{2}(\partial u_{i}^{(s)}/{\partial x_j}+{\partial u_{j}^{(s)}}/{\partial x_i})$. The relation of displacement operator $\vec u^{(s)}$ and $\vec u_{\mu}(\vec k)$ is set up by Fourier transformation:
\begin{eqnarray}\label{A3}
u_{i}^{(s)}=\frac{1}{\sqrt{N}}\sum_{\vec k\mu}u_{\mu}(\vec k){\rm e}_{\mu i}(\vec k)e^{i\vec k\cdot\vec x_s}
\end{eqnarray}
where $\vec {\rm e}_{\mu }(\vec k)$ is the unit vector representing the direction of vibrations, $N$ is the number of particles in an elementary cell of the glass sample. For longitudinal phonon mode with $\mu=l$,
$
{\rm e}_{li}(\vec k)={k_i}/{|\vec k|}
$, whereas for transverse modes with $\mu=t_1$ and $t_2$, we have, 
\begin{eqnarray}\label{A4}
 & {} & \vec {\rm e}_{t_1}(\vec k)\cdot\vec k=\vec {\rm e}_{t_2}(\vec k)\cdot \vec k=\vec {\rm e}_{t_1}(\vec k)\cdot \vec {\rm e}_{t_1}(\vec k)=0\nonumber \\
 & {} & \sum_{\mu=t_1,t_2}{\rm e}_{\mu i}(\vec k){\rm e}_{\mu j}(\vec k)=\delta_{ij}-\frac{k_ik_j}{k^2}
\end{eqnarray}
the strain field is therefore expressed as 
$
e_{ij}^{(s)}=\frac{1}{2\sqrt{N}}\sum_{\vec k\mu}iu_{\mu}(\vec k)e^{i\vec k\cdot \vec x_s}[k_j{\rm e}_{\mu i}(\vec k)+k_{i}{\rm e}_{\mu j}(\vec k)]
$. For an arbitrary function $f(\vec k)$, we always have the following relation, 
$
\sum_{\vec k}f(\vec k)=\sum_{\vec k}\frac{1}{2}[f(\vec k)+f(-\vec k)]
$. The displacement operator $u_{i}^{(s)}$ is a real quantity, i.e., $u_{i}^{(s)}=u_i^{(s)*}$, we have $
u_{\mu i}(\vec k)=u_{\mu i}^*(-\vec k)
$.
Finally, with the properties of $u_{\mu}(\vec k)$ operator, we rewrite the stress-strain coupling as follows, 
\begin{eqnarray}\label{A5}
\sum_s\sum_{ij}e_{ij}^{(s)}\hat{T}_{ij}^{(s)}
 & = & \frac{1}{4\sqrt{N}}\sum_{ij}\sum_s\sum_{\vec k\mu}\left[\left(iu_{\mu}(\vec k)e^{i\vec k\cdot \vec x_s}\right)+\left(iu_{\mu}(\vec k)e^{i\vec k\cdot \vec x_s}\right)^*\right](k_j{\rm e}_{\mu j}(\vec k)+k_j{\rm e}_{\mu i}(\vec k))\hat{T}_{ij}^{(s)}
\end{eqnarray}
Because the stress-strain coupling is linear in the displacement operator $u_{\mu}(\vec k)$, we can absorb it into the term which is quadratic in $u_{\mu}(\vec k)$, by ``completing the square". 
\begin{eqnarray}\label{A6}
\hat{H}=\sum_{\vec k\mu}\left(\frac{|p_{\mu}(\vec k)|^2}{2m}+\frac{m\omega^2_{\vec k\mu}}{2}|u_{\mu}(\vec k)-u_{\mu}^{(0)}(\vec k)|^2-\frac{m\omega^2_{\vec k\mu}}{2}|u_{\mu}^{(0)}(\vec k)|^2\right)+\hat{H}^{\rm non}
\end{eqnarray}
where the ``equilibrium position" $u_{\mu}^{(0)}(\vec k)$ is given as follows
\begin{eqnarray}\label{A7}
u_{\mu}^{(0)}(\vec k)=\frac{i}{2\sqrt{N}m\omega_{\vec k\mu}^2}\sum_{ij}\sum_s\bigg[k_j{\rm e}_{\mu i}(\vec k)+k_{i}{\rm e}_{\mu j}(\vec k)\bigg]\hat{T}_{ij}^{(s)}e^{-i\vec k\cdot \vec x_s}
\end{eqnarray}
The extra term left out after completing the square is the effective interaction between non-elastic stress tensors. It can be rewritten into two parts, the first part represents non-elastic stress-stress interaction within the single block, while the second part represents the interaction between different blocks:
\begin{eqnarray}\label{A8}
 & {} & -\sum_{\vec k\mu}\left(\frac{m\omega^2_{\vec k\mu}}{2}|u_{\mu}^{(0)}(\vec k)|^2\right)\nonumber \\
 & = & -\sum_{\vec k\mu}\frac{1}{8Nm\omega^2_{\vec k\mu}}\sum_{ijkl}\bigg[k_{j}{\rm e}_{\mu i}(\vec k)+k_i{\rm e}_{\mu j}(\vec k)\bigg]
\bigg[k_{k}{\rm e}_{\mu l}(\vec k)+k_l{\rm e}_{\mu k}(\vec k)\bigg]\sum_s\hat{T}_{ij}^{(s)}\hat{T}_{kl}^{(s')}\nonumber \\
 & {} & -\sum_{\vec k\mu}\frac{1}{8Nm\omega^2_{\vec k\mu}}\sum_{ijkl}\bigg[k_{j}{\rm e}_{\mu i}(\vec k)+k_i{\rm e}_{\mu j}(\vec k)\bigg]
\bigg[k_{k}{\rm e}_{\mu l}(\vec k)+k_l{\rm e}_{\mu k}(\vec k)\bigg]\sum_{s\neq s'}\hat{T}_{ij}^{(s)}\hat{T}_{kl}^{(s')} \cos(\vec k\cdot (\vec x_s-\vec x'_s))
\end{eqnarray}
We denote the second term in Eq.(\ref{A8}) as $\hat{V}$, the non-elastic stress-stress interaction. Applying the properties of unit vector for longitudinal and transverse phonons, it is further simplified as 
\begin{eqnarray}\label{A9}
 & {} & \hat{V} 
= \sum_{ijkl}\sum_{s\neq s'}\Lambda_{ijkl}^{(ss')}\hat{T}_{ij}^{(s)}\hat{T}_{kl}^{(s')}\nonumber \\
 & {} & \Lambda_{ijkl}^{(ss')}= \frac{1}{a^3}\sum_{\vec k}e^{i\vec k\cdot (\vec x_s-\vec x_s')}\Lambda_{ijkl}(\vec k)\nonumber \\
 & {} & \Lambda_{ijkl}(\vec k) = \frac{1}{2\rho}\left(\frac{1}{c_t^2}-\frac{1}{c_l^2}\right)\left(\frac{k_ik_jk_kk_l}{k^4}\right)-\frac{1}{8\rho c_t^2}\left(\frac{k_jk_l\delta_{ik}+k_jk_k\delta_{il}
+k_ik_l\delta_{jk}+k_ik_k\delta_{jl}}{k^2}\right)
\end{eqnarray}
where $\rho =Nm/a^3$, with $a$ the length scale of the elementary block of glass. Eq.(\ref{A8}) is the non-elastic stress-stress interaction without the presence of external static strain field. If we plug in the external static strain $\bm{e}$, then the non-elastic stress-stress interaction $\hat{V}(\bm{e})$ is given as follows, 
\begin{eqnarray}\label{A10}
 & {} & \hat{V}(\bm{e}) 
= \sum_{ijkl}\sum_{ss'}\Lambda_{ijkl}^{(ss')}(\bm{e})\hat{T}_{ij}^{(s)}(\bm{e})\hat{T}_{kl}^{(s')}(\bm{e})\nonumber \\
 & {} & \Lambda_{ijkl}^{(ss')}(\bm{e}) = \frac{1}{(1+e_{xx})a^3}\sum_{\vec k}e^{i\vec k\cdot (\vec x_s+\vec u^{(s)}-\vec x'_s-\vec u^{(s')})}\Lambda_{ijkl}(\vec k)\nonumber \\
 & {} & \Lambda_{ijkl}(\vec k) = \frac{1}{2\rho}\left(\frac{1}{c_t^2}-\frac{1}{c_l^2}\right)\left(\frac{k_ik_jk_kk_l}{k^4}\right)-\frac{1}{8\rho c_t^2}\left(\frac{k_jk_l\delta_{ik}+k_jk_k\delta_{il}
+k_ik_l\delta_{jk}+k_ik_k\delta_{jl}}{k^2}\right)
\end{eqnarray}
where we use $\vec u^{(s)}$ to denote the displacement of a certain particle located at position $\vec x_s$ which is driven by external static, uniform strain $\bm{e}=e_{xx}$. $(1+e_{xx})a^3$ is the volume of the elementary block under the deformation of external static strain.

In this section we want to give a detailed discussion about the coefficient of non-elastic stress-stress interaction which appears in Eq.(\ref{23}): $\Lambda_{ijkl}^{(ss')}(\bm{e})$. 
The super block length in the $n$-th step renormalization is $N_0L_n$, with the length $(1+e_{xx})N_0L_n$, width $N_0L_n$ and height $N_0L_n$. Therefore, we always have $|\vec x_s-\vec x_s'|\le (1+e_{xx})N_0L_{n}$ for arbitrary blocks at positions $\vec x_s$ and $\vec x_s'$ (that is, the distance between single blocks within a super block must be no greater than the length of the super block $(1+e_{xx})N_0L_n$). Let us write the coefficient $\Lambda_{ijkl}^{(ss')}(\bm{e})$ into two parts:
\begin{eqnarray}\label{A1}
\Lambda_{ijkl}^{(ss')}(\bm{e}) & = & \frac{1}{(1+e_{xx})a^3}\left(\sum_{|k_{x}|=\frac{2\pi}{(1+e_{xx})R}}^{\frac{2\pi}{(1+e_{xx})N_0L_n}}\sum_{|k_y|, |k_z|=\frac{2\pi}{R}}^{\frac{2\pi}{N_0L_n}}+\sum_{| k_x|=\frac{2\pi}{(1+e_{xx})N_0L_n}}^{\frac{2\pi}{(1+e_{xx})L_1}}\sum_{|k_y|, |k_z|=\frac{2\pi}{N_0L_n}}^{\frac{2\pi}{L_1}}\right)\Lambda_{ijkl}(\vec k)e^{i\vec k\cdot (\vec x_s+\vec u^{(s)}-\vec x'_s-\vec u^{(s')})}\nonumber \\
\end{eqnarray}
where $R$ is the experimental length scale. Therefore, the first part of the above summation, can be simplified as $\sum_{|\vec k|=2\pi/R}^{2\pi/N_0L_n}\Lambda_{ijkl}(\vec k)e^{i\vec k\cdot (\vec x_s+\vec u^{(s)}-\vec x_s'-\vec u^{(s')})}\approx \sum_{|\vec k|=2\pi/R}^{2\pi/N_0L_n}\Lambda_{ijkl}(\vec k)$.
\begin{eqnarray}\label{A2}
\Lambda_{ijkl}^{(ss')}(\bm{e}) & = & \frac{1}{(1+e_{xx})a^3}\left(\sum_{|k_{x}|=\frac{2\pi}{(1+e_{xx})R}}^{\frac{2\pi}{(1+e_{xx})N_0L_n}}\sum_{|k_y|, |k_z|=\frac{2\pi}{R}}^{\frac{2\pi}{N_0L_n}}\Lambda_{ijkl}(\vec k)+\sum_{| k_x|=\frac{2\pi}{(1+e_{xx})N_0L_n}}^{\frac{2\pi}{(1+e_{xx})L_1}}\sum_{|k_y|, |k_z|=\frac{2\pi}{N_0L_n}}^{\frac{2\pi}{L_1}}\Lambda_{ijkl}(\vec k)e^{i\vec k\cdot (\vec x_s+\vec u^{(s)}-\vec x'_s-\vec u^{(s')})}\right)\nonumber \\
\end{eqnarray}

After summing over different directions of momentum $\vec k$, the first part of Eq.(\ref{A2}) is given by 
\begin{eqnarray}\label{A3}
 & {} & \frac{1}{(1+e_{xx})a^3}\left(\sum_{|k_{x}|=\frac{2\pi}{(1+e_{xx})R}}^{\frac{2\pi}{(1+e_{xx})N_0L_n}}\sum_{|k_y|, |k_z|=\frac{2\pi}{R}}^{\frac{2\pi}{N_0L_n}}\Lambda_{ijkl}(\vec k)\right)\nonumber \\
 & {} & =\frac{4\pi}{3}\frac{1}{(1+e_{xx})(N_0L_n)^3}\left[\frac{\alpha}{30\rho c_t^2}\left(\delta_{ij}\delta_{kl}+\delta_{ik}\delta_{jl}+\delta_{il}\delta_{jk}\right)-\frac{1}{4\rho c_t^2}\left(\delta_{jl}\delta_{ik}+\delta_{jk}\delta_{il}\right)\right]
\end{eqnarray}
where $\alpha=1-c_t^2/c_l^2$. The second term of Eq.(\ref{A2}) is obtained by D. Zhou and A. J. Leggett\cite{Zhou2015-1}:
\begin{eqnarray}\label{A4}
 & {} & \frac{1}{(1+e_{xx})a^3}\left(\sum_{| k_x|=\frac{2\pi}{(1+e_{xx})N_0L_n}}^{\frac{2\pi}{(1+e_{xx})L_1}}\sum_{|k_y|, |k_z|=\frac{2\pi}{N_0L_n}}^{\frac{2\pi}{L_1}}\Lambda_{ijkl}(\vec k)e^{i\vec k\cdot (\vec x_s+\vec u^{(s)}-\vec x'_s-\vec u^{(s')})}\right)
= -\frac{\tilde{\Lambda}_{ijkl}}{8\pi\rho c_t^2|\vec x_s+\vec u^{(s)}-\vec x'_s-\vec u^{(s')}|^3}\nonumber \\
 & {} & \tilde{\Lambda}_{ijkl}=\frac{1}{4}\bigg\{(\delta_{jl}-3n_jn_l)\delta_{ik}+(\delta_{jk}-3n_jn_k)\delta_{il}+(\delta_{ik}-3n_in_k)\delta_{jl}
+(\delta_{il}-3n_in_l)\delta_{jk}\bigg\}\nonumber \\
 & {} & +\frac{1}{2}\alpha\bigg\{-(\delta_{ij}\delta_{kl}+\delta_{ik}\delta_{jl}+\delta_{jk}\delta_{il})+3(n_in_j\delta_{kl}+n_in_k\delta_{jl}+n_in_l\delta_{jk}+n_jn_k\delta_{il}+n_jn_l\delta_{ik}+n_kn_l\delta_{ij})-15n_in_jn_kn_l\bigg\}\nonumber \\
\end{eqnarray}
where $\vec n$ is the unit vector of $\vec x_s+\vec u^{(s)}-\vec x'_s-\vec u^{(s')}$. Finally, the coefficient $\Lambda_{ijkl}^{(ss')}(\bm{e})$ which appears in Eq.(\ref{23}) is the summation of Eq.(\ref{A3}) and Eq.(\ref{A4}).

\endwidetext

\end{document}